\documentclass[a4paper,11pt]{article} 
\usepackage{jcappub} 
\pdfoutput=1 
\usepackage[utf8]{inputenc} 
\usepackage{subcaption} 
\usepackage{threeparttable} 
\usepackage{multirow}
\usepackage[table]{xcolor} 
\usepackage[noabbrev]{cleveref} 
\Crefname{equation}{Eq.}{Eqs.} 
\usepackage{aasmacros} 
\usepackage[normalem]{ulem} 

\title{\boldmath{J-PAS: forecast on the primordial power spectrum reconstruction}}

\author[a,b]{Guillermo Martínez-Somonte,}
\author[a]{Airam Marcos-Caballero,}
\author[a]{Enrique Martínez-González,}
\author[c]{Antonio L. Maroto,}
\author[d,e,f]{Miguel Quartin,}
\author[g]{Raul Abramo,}
\author[h]{Jailson Alcaniz,}
\author[i]{Narciso Benítez,}
\author[j,k]{Silvia Bonoli,}
\author[h]{Saulo Carneiro,}  
\author[k,l]{Javier Cenarro,} 
\author[k]{David Cristóbal-Hornillos,}
\author[h]{Simone Daflon,}
\author[h]{Renato Dupke,}
\author[k,l]{Alessandro Ederoclite,} 
\author[m]{Rosa María González Delgado,}
\author[k,l]{Antonio Hernán-Caballero,}
\author[n,o]{Carlos Hernández–Monteagudo,}
\author[p]{Jifeng Liu,}
\author[k,l]{Carlos López-Sanjuan,}
\author[k,l]{Antonio Marín-Franch,}
\author[q]{Claudia Mendes de Oliveira,}
\author[k]{Mariano Moles,}
\author[h]{Fernando Roig,}
\author[q]{Laerte Sodré Jr.,}  
\author[r]{Keith Taylor,}
\author[k]{Jesús Varela,} 
\author[k,l]{Héctor Vázquez Ramió,} 
\author[m]{José M. Vilchez}
\author[k,l]{and Javier Zaragoza-Cardiel}

\affiliation[a]{Instituto de Física de Cantabria, CSIC-Universidad de Cantabria, Avenida de los Castros s/n, E-39005 Santander, Spain}
\affiliation[b]{Departamento de Física Moderna, Universidad de Cantabria, Avenida de los Castros s/n, E-39005 Santander, Spain}
\affiliation[c]{Departamento de Física Teórica and Instituto de Física de Partículas y del Cosmos (IPARCOS-UCM),  Universidad Complutense de Madrid,\\28040 Madrid, Spain}
\affiliation[d]{Centro Brasileiro de Pesquisas Físicas, 22290-180, Rio de Janeiro, RJ, Brazil}
\affiliation[e]{Observatório do Valongo, Universidade Federal do Rio de Janeiro, 20080-090, Rio de Janeiro, RJ,
Brazil}
\affiliation[f]{PPG Cosmo, Universidade Federal do Espírito Santo, 29075-910, Vitória, ES, Brazil}
\affiliation[g]{Departamento de Física Matemática, Instituto de Física, Universidade de São Paulo, Rua do Matão 1371, 05508-090, São Paulo, SP, Brazil}
\affiliation[h]{Observatório Nacional, Rua General José Cristino, 77, São Cristóvão, 20921-400, Rio de Janeiro, RJ, Brazil}
\affiliation[i]{Independent Researcher}
\affiliation[j]{Donostia International Physics Center (DIPC), Manuel Lardizabal Ibilbidea, 4, San Sebastián, Spain}
\affiliation[k]{Centro de Estudios de Física del Cosmos de Aragón (CEFCA), Plaza San Juan, 1, E-44001, Teruel, Spain}
\affiliation[l]{Unidad Asociada CEFCA-IAA, CEFCA, Unidad Asociada al CSIC por el IAA y el IFCA, Plaza San Juan 1, 44001 Teruel, Spain}
\affiliation[m]{Instituto de Astrofísica de Andalucía (IAA) - CSIC, Apdo 3004, E-18080, Granada, Spain}
\affiliation[n]{Instituto de Astrofísica de Canarias, C/ Vía Láctea, s/n, E-38205, La Laguna, Tenerife, Spain}
\affiliation[o]{Universidad de La Laguna, Avda. Francisco Sánchez, E-38206, San Cristóbal de La Laguna, Tenerife, Spain}
\affiliation[p]{Chinese National Astronomical Observatory of China,  Chinese Academy of Sciences, Beijing, China}
\affiliation[q]{Departamento de Astronomia, Instituto de Astronomia, Geofísica e Ciências Atmosféricas, Universidade de São Paulo, São Paulo, Brazil}
\affiliation[r]{Instruments4, 4121 Pembury Place, La Canada Flintridge, CA 91011, U.S.A.}

\emailAdd{gmsomonte@ifca.unican.es}
\emailAdd{marcos@ifca.unican.es}
\emailAdd{martinez@ifca.unican.es}
\emailAdd{maroto@ucm.es}
\emailAdd{mquartin@if.ufrj.br}
\emailAdd{abramo@fma.if.usp.br}
\emailAdd{alcaniz@on.br}
\emailAdd{txitxo.benitez@gmail.com}
\emailAdd{silvia.bonoli@dipc.org}
\emailAdd{saulocarneiro@on.br}
\emailAdd{cenarro@cefca.es}
\emailAdd{dchornillos@gmail.com}
\emailAdd{daflon@on.br}
\emailAdd{rdupke@gmail.com}
\emailAdd{aederocl.astro@gmail.com}
\emailAdd{rosa@iaa.es}
\emailAdd{ahernan@cefca.es}
\emailAdd{chm@iac.es}
\emailAdd{jfliu@nao.cas.cn}
\emailAdd{clsj@cefca.es}
\emailAdd{amarin@cefca.es}
\emailAdd{claudia.oliveira@iag.usp.br}
\emailAdd{moles@cefca.es}
\emailAdd{froig@on.br}
\emailAdd{laerte.sodre@iag.usp.br}
\emailAdd{kt.astro@gmail.com}
\emailAdd{jvarela@cefca.es}
\emailAdd{hvr@cefca.es}
\emailAdd{jvm@iaa.csic.es}
\emailAdd{jzaragoza@cefca.es}

\abstract{We investigate the capability of the J-PAS survey to constrain the primordial power spectrum using a non-parametric Bayesian method. Specifically, we analyze simulated power spectra generated by an oscillatory primordial feature template motivated by non-standard inflation. The feature is placed within the range of scales where the signal-to-noise ratio is maximized, and we restrict the analysis to $k \in [0.02,0.2] \text{ h} \text{ Mpc}^{-1}$, set by the expected J-PAS coverage and the onset of non-linear effects. Each primordial power spectrum is reconstructed by linearly interpolating $N$ knots in the $\{\log k, \log P_{\mathcal{R}}(k)\}$ plane, which are sampled jointly with the cosmological parameters $\{H_0,\Omega_b h^2, \Omega_c h^2\}$ using PolyChord. To test the primordial features, we apply two statistical tools: the Bayes factor and a hypothesis test that localizes the scales where features are detected. We assess the recovery under different J-PAS specifications, including redshift binning, tracer type, survey area, and filter strategy. Our results show that combining redshift bins and tracers allows the detection of oscillatory features as small as 2\%.}

\keywords{Inflation, Bayesian inference, galaxy surveys, power spectrum}

\begin{document}

\maketitle
\flushbottom 

\section{Introduction}
\label{sec:Introduction}

Forthcoming galaxy surveys represent the next frontier in exploring the initial conditions of the Universe. Cosmological primordial fluctuations have been determined to be adiabatic, Gaussian, and quasi-scale invariant by various cosmological observations, as well as a background universe spatially isotropic and homogeneous \cite{Planck18Parameters}. Multiple shortcomings of the standard Hot Big Bang scenario \cite{InflaGuth1981,InflaLinde1982}, together with these properties, strongly motivate cosmological inflation \cite{InflaGuth1981,InflaLinde1982,InflaBrout1978,InflaStarobinski1980,InflaAlbrechtSteinhardt1982,InflaLinde1983}, a hypothetical epoch of exponential expansion in the early universe whose nature and origin remain uncertain and loosely constrained by current observations.

Inflationary models can be constrained by measuring the primordial power spectrum of curvature perturbations, $P_\mathcal{R}(k)$, which encodes valuable information about the physical mechanism responsible for generating the initial conditions of cosmic structure formation. $P_\mathcal{R}(k)$ is usually parametrized by a simple power law with two parameters: the amplitude $A_s$ and the spectral index $n_s$ of the primordial comoving curvature perturbations. The latest results from the ESA Planck satellite provide a value for the scalar power spectrum parameters of $A_s = \left( 2.10_{-0.04}^{+0.03} \right) \times 10^{-9}$  and $n_s = 0.965 \pm 0.004$ \cite{Planck18Parameters}. The predictions of simple single field slow-roll models of inflation \cite{Starobinski1979, InflaLinde1982} are consistent with the latest Planck results, and models such as Higgs inflation \cite{HiggsInflation} or Starobinsky $R^2$ inflation \cite{StarobinskiInflation} provide good fits to the latest Planck plus BICEP2/Keck Array data \cite{PlanckInflation18, BICEPKeck, JudgmentDayInflation}. These results are compatible with a wide range of slow-roll inflationary models, many of which differ in their field content, potential shapes, and predictions for primordial observables \cite{EnciclopediaInflation}.

Departures from the slow-roll scenario can lead to the production of distinctive primordial features in $P_\mathcal{R}(k)$. Therefore, it is important to identify these observational signatures, which may arise either from departures in slow-roll dynamics \cite{MirandaHuModeloFeature,SearchingFeatures1,SearchingFeatures2} or from modified scalar field dynamics during inflation \cite{InverseScalarDynamics}. Common deviations from the power-law form of $P_{\mathcal{R}}(k)$ are: global logarithmic oscillations arising from non-Bunch-Davies initial conditions \cite{ModelMartin2001,ModelMartin2003,ModelBozza2003} or from axion monodromy \cite{ModelFlauger2017}; global linear oscillations predicted by boundary effective field theory models \cite{SearchingFeatures1, ModelJackson2013} and localized oscillatory features induced by a step in the inflaton potential \cite{ModelAdams2001} or in the sound speed \cite{ModelAchucarro2010, ModelInflationSpeedSound}. Also models with large scales cutoffs \cite{ModelContaldi2003,Parametric2Sinha2006} or more general modulations \cite{ModelDanielsson2002,ModelChen2012} can be found in the literature.

Stage IV galaxy surveys \cite{JPASSpecifications,DESISpecifications,EuclidSpecifications} are expected to provide significantly improved constraints on primordial features compared to Cosmic Microwave Background (CMB) experiments at intermediate scales $k \sim 0.01-0.5 \text{ h} \text{ Mpc}^{-1}$. The surveys are typically classified as either spectroscopic or photometric. Spectroscopic surveys measure the full spectrum of light from astronomical objects, enabling precise determinations of redshift, chemical composition, and line-of-sight velocity through detailed spectral features.
Examples include the Baryon Oscillation Spectroscopic Survey (BOSS) \cite{BOSSGeneral1,BOSSGeneral2}, the extended Baryon Oscillation Spectroscopic Survey (eBOSS) \cite{eBOSS1,eBOSS2}, or the Dark Energy Spectroscopic Instrument (DESI) \cite{DESISpecifications}. While these surveys provide high-resolution redshift measurements, they are limited in object number due to the time-intensive nature of spectroscopic observations. In contrast, photometric surveys such as the Dark Energy Survey (DES) \cite{DESGeneric1,DESGeneric2} and the Hyper Suprime-Cam Subaru Strategic Program (HSC-SSP) \cite{HSCSSP} obtain broadband photometry without spectral dispersion, allowing for the observation of vastly larger samples at the cost of lower redshift precision. Other LSS surveys use a combination of photometry and spectroscopy. This is the case of Euclid \cite{EuclidSpecifications}, which employs traditional broad-band photometry and slitless spectroscopy, as well as J-PAS (Javalambre Physics of the Accelerated Universe Astrophysical Survey) \cite{Benitez2009, JPASSpecifications}, which primarily relies on high-resolution photometric techniques using a large number of narrow-band filters to achieve spectroscopic-like precision. The high signal-to-noise ratio $S/N$ achievable in the galaxy power spectra of these surveys---depending on factors such as galaxy number density, redshift uncertainty, and galaxy bias---enables sensitive searches for physically motivated features imprinted in the primordial power spectrum.

The shape of the primordial power spectrum has been determined using two different approaches: parametrizations and reconstructions\footnote{Other approaches, such as the \texttt{FreePower} method \cite{FreePower}, can infer the shape of the primordial power spectrum indirectly by reconstructing the linear matter power spectrum.}. Reconstructions of $P_\mathcal{R}(k)$ derive its shape directly from observational data without imposing any predefined model or template, in contrast to parametric approaches \cite{ModelContaldi2003,ParametricBridle, Parametric3Simon2005,Parametric4Bridges2006,Parametric2Sinha2006,Parametric6Covi2006,Parametric5Bridges2007,Parametric7Joy2009,Parametric8Paykari2010,Parametric9Guo2011,Parametric10Goswami2013, Ballardini1, Ballardini2, Ballardini3, EuclidSearchFeatures, PrimordialFeaturesEFT2025,SimonySantosParametric}. Several methods have been developed for model-independent reconstructions of $P_\mathcal{R}(k)$ based on different inference approaches \cite{BallardiniCMB2025, ReconLinearInterpolation1,ReconTopHat, ReconWaveletReal, ReconWaveletSimulated, ReconSplines, ReconSplinesWMAP1, ReconSplinesWMAP3, PlanckInflation13, PlanckInflation15, PlanckInflation18, ReconFilter, ReconPRISM, ReconLSSOrthogonal, WillPaper,KnotedSky, ReconCMBlike}. The precision of these methods varies from subpercent to $30\%$, and no statistically significant deviations from the primordial power-law have been found using galaxy survey data \cite{ReconTopHat,ReconPRISM,ReconSplines,ReconSplinesWMAP3}. In addition, reconstructions using CMB data exclusively, such as the Planck mission, have also shown consistency with the Standard Model \cite{PlanckInflation13,PlanckInflation15,PlanckInflation18,ReconLinearInterpolation1,KnotedSky,ReconSplinesWMAP1,ReconCMBlike,WillPaper,ReconstructionsCMB2025}.

Following the reconstruction approach, coarse features encompassing multiple scales in $P_{\mathcal{R}}(k)$ can be identified, but fine ones can be difficult to detect \cite{FineFeatures}. In contrast, parametric approaches can detect fine features more easily, but strongly depend on a previously assumed model or template of $P_{\mathcal{R}}(k)$. Justifying which template to select may be challenging due to the vast number of proposed models. The reconstructions are more appropriate in order to obtain model-independent information.
 
The methodology used in this work consists in reconstructing $P_{\mathcal{R}}(k)$ sampling the placement of pairs of points in the log $\{k,P_{\mathcal{R}}\}$ plane, named `knots' \cite{WillPaper,MethodologicalPaper}. We determine the $P_{\mathcal{R}}(k)$ by performing a linear interpolation of these knots covering all the scales of interest. This method does not require any prior $k$-binning of $P_{\mathcal{R}}(k)$, thus allowing to detect features and to locate them at any scales. We compute the Bayes factor to quantify how the highest-evidence $N$-knot reconstruction of $P_{\mathcal{R}}(k)$ compares to the power-law spectrum, which corresponds to the $N = 2$ configuration. Our methodology is flexible and adaptable to data from diverse surveys. A limitation of this approach, however, lies in the lack of smoothness introduced by the linear splines connecting the knots, which other methods palliate by using cubic splines or other smoothing techniques. Our method may also struggle with very sharp features or features with many oscillations, which is a common drawback of non-parametric approaches.

Our objective is to forecast the capability of J-PAS galaxy clustering to detect features in the primordial power spectrum $P_\mathcal{R}(k)$, using a non-parametric Bayesian method \cite{MethodologicalPaper} that does not assume any specific inflationary model. We consider different observational configurations and specifications, as redshift bin, tracer type, area of the sky, or tray strategy for the filters, and explore different simulated features, in order to find out which feature amplitudes, and under which configurations and survey specifications, could be detected with J-PAS.

The structure of the paper is as follows: \cref{sec:J-PAS} provides a brief overview of J-PAS, including the survey specifications used in this work. In \cref{sec:PPSReconstructions} we detail the procedure to reconstruct $P_\mathcal{R}(k)$ with knots through nested sampling, along with the details of the tests applied for the subsequent analysis. \Cref{sec:GPSModel} describes the galaxy power spectrum model that we use for J-PAS, and in \cref{sec:SNSensitivity} we give an analysis of its sensitivity. In \cref{sec:Results} we explore how a local oscillatory feature can be detected depending on the J-PAS tracer, area, and tray strategy. We also combine multiple galaxy clustering information to explore the minimum detectable amplitude. \Cref{sec:Conclusions} presents the conclusions derived from this work and some possible lines of future work. In the Appendix we explore the impact of the BAO smoothing on the reconstructions.

\section{The J-PAS survey}
\label{sec:J-PAS}

The Javalambre Physics of the Accelerated Universe Astrophysical Survey (J-PAS) \cite{Benitez2009, JPASSpecifications} is an astronomical project designed to map the Large Scale Structure (LSS) of the universe and study the nature of dark energy by observing millions of galaxies over thousands of square degrees in the Northern Hemisphere. Conducted with the 2.5 meter Javalambre Survey Telescope (JST) at Sierra de Javalambre, in Teruel, Spain, the instrument that makes J-PAS possible is JPCam, which has 14 CCDs, each one containing 9200×9200 pixels, and covering an area of $\sim 4.2 \text{ deg}^2$ in a single pointing. J-PAS employs a multi-filter system featuring 57 optical filters. The combination of 54 contiguous narrow-band filters provides an effective spectral resolution of $R$ $\sim$ 60 across the 3800–9100 $\text{\r{A}}$ range, delivering low-resolution spectrophotometry for each $0''.48 \times 0''.48$ pixel on the sky. This allows the survey to capture precise redshifts and spectral energy distributions for diverse celestial objects, including galaxies, stars, and quasars (QSOs). The spectro-photometric approach of J-PAS combines the efficiency of photometric surveys with the spectral detail of spectroscopic ones, making it a powerful tool for cosmology and galaxy evolution studies.
The unique filter configuration allows for an accurate photometric error determination, ensuring a high signal-to-noise ratio to resolve fine spectral features. This approach enables J-PAS to classify objects with high accuracy, identifying Emission-Line Galaxies (ELGs), Luminous Red Galaxies (LRGs), faint dwarf galaxies, or stellar populations, while also providing insights into star formation history, active galactic nuclei, and cosmological parameters like the growth rate of structures. Prior to the JPCam's installation, the JST telescope used a single-CCD camera to conduct the miniJPAS survey \cite{MiniJPAS}, a $\sim 1 \text{ deg}^2$ precursor using the full J-PAS filter set to test algorithms \cite{MiniJPASEmissionLines1, MiniJPASEmissionLines2}, classify objects \cite{MiniJPASClassification1, MiniJPASClassification2, MiniJPASClassificationQSO}, study galaxies \cite{MiniJPASGalaxies1, MiniJPASGalaxies2, MiniJPASGalaxies3, MiniJPASGalaxies4, MiniJPASGalaxiesEvolution} and QSOs \cite{MiniJPASQSOs}, and estimate photo-z and object densities \cite{MiniJPASPhotoZ1, MiniJPASPhotoZ2}.

J-PAS has been taking data since October 2023, and this work presents a forecast for the full survey. Several J-PAS forecasts have appeared in the literature, addressing a variety of cosmological topics. \cite{JPASForecastNeutrinos2} constrains the sum of neutrino masses in combination with CMB and/or supernova data, reaching $\Sigma m_{\nu} < 0.061\text{ eV}$. \cite{JPASForecastNeutrinos1} considers a time-varying dark energy equation of state and obtains a tighter constraint of $\Sigma m_{\nu} < 0.017\text{ eV}$ when combining J-PAS with LSS and CMB data. Both studies highlight the potential of J-PAS to distinguish between the normal and inverted neutrino mass hierarchy. \cite{JPASForecastVacuumEnergy} provides parameter constraints for a class of interacting vacuum energy models, in some cases more precise than those expected from DESI and Euclid in the low-redshift range. \cite{JPASForecastCouplingDarkEnergyDarkMatter} examines the capability of J-PAS to distinguish between coupled and uncoupled dark matter--dark energy models; by measuring galaxy clustering suppression, they find that J-PAS data could detect an interaction at the few-percent level with a significance greater than $10\sigma$. \cite{JPASForecastDarKEnergyModifyGravity} constrains deviations from General Relativity by measuring the effective Newton constant and the gravitational slip parameter with 2–7\% precision, surpassing the expected precision of DESI and Euclid forecasts in the redshift range $z = 0.3-0.6$. That work also constrains the time-dependent dark energy equation-of-state parameters, reaching absolute errors of $\delta \omega_0 = 0.058$ and $\delta \omega_a = 0.24$. Finally, \cite{JPASForecastDarkEnergyCosta} forecasts the ability of J-PAS to constrain parameters of interacting dark energy models, finding an absolute uncertainty in the interaction parameter of about 0.02. Collectively, these studies demonstrate the strong potential of J-PAS for cosmology, in many cases providing forecasts comparable to or exceeding those from DESI and Euclid.

To perform this forecast, we use 
photometric redshifts (photo-$z$) and galaxy number densities estimated from the 2024 Internal Data Release (IDR) of J-PAS, 
based on observations covering a sky area of $27 \text{ deg}^2$. These estimates are broadly consistent with those from miniJPAS, although some differences arise in specific redshift bins. Using the IDR-based estimates, we simulate the J-PAS galaxy power spectrum across the full expected survey area to forecast the detectability of primordial features. Our analysis includes both blue and red galaxy populations, spanning redshift bins from $z=0.1$ to $z=1.1$ with bin width $\Delta z=0.1$.

The J-PAS 57 optical filters (54 narrow-band, 2 medium-band, and 1 broad-band) are distributed in different mechanical trays. The tray configuration is designed to group filters within a specific wavelength range, placing some at the edges and others at the center, in order to minimize image artifacts such as `ghosts' or reflections. Trays 1 and 2 (T1 and T2) primarily contain blue filters, except in the central CCDs, where red filters are used precisely to reduce reflections. The opposite configuration is applied to trays 3 and 4 (T3 and T4). T5 contains filters covering the Sloan Digital Sky Survey $i$-band. Various observational strategies can be followed, with designs tailored to balance area coverage and efficiency in capturing photometric data. Ideally, the highest photo-$z$ qualities are obtained from the entire tray system of J-PAS (T12345) employing all the 57 filters. In practice, we observe that using a tray strategy with the bluest trays and T5 only (T125) results in photo-$z$ errors that are within a few tens of percent of those obtained from the full tray system T12345, while enabling coverage of more than twice the sky area in the same observing time. The photo-$z$ errors tend to increase with $z$, in contrast to the galaxy densities: while both strategies provide comparable densities at low $z$-bins, differences of up to an order of magnitude are found at higher redshifts.
In this work, we compare the performance of these two observational strategies for reconstructing $P_\mathcal{R}(k)$. 

\begin{figure}[t]
\centering 
\includegraphics[width=.495\textwidth]{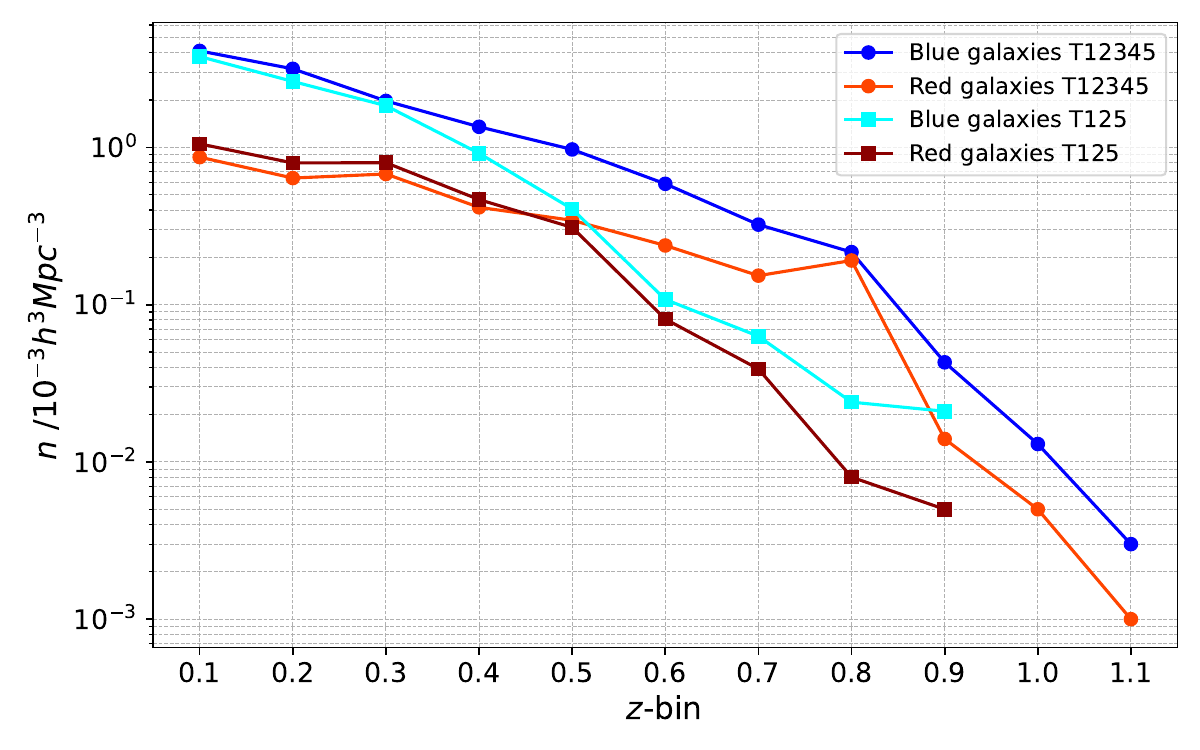}
\includegraphics[width=.495\textwidth]{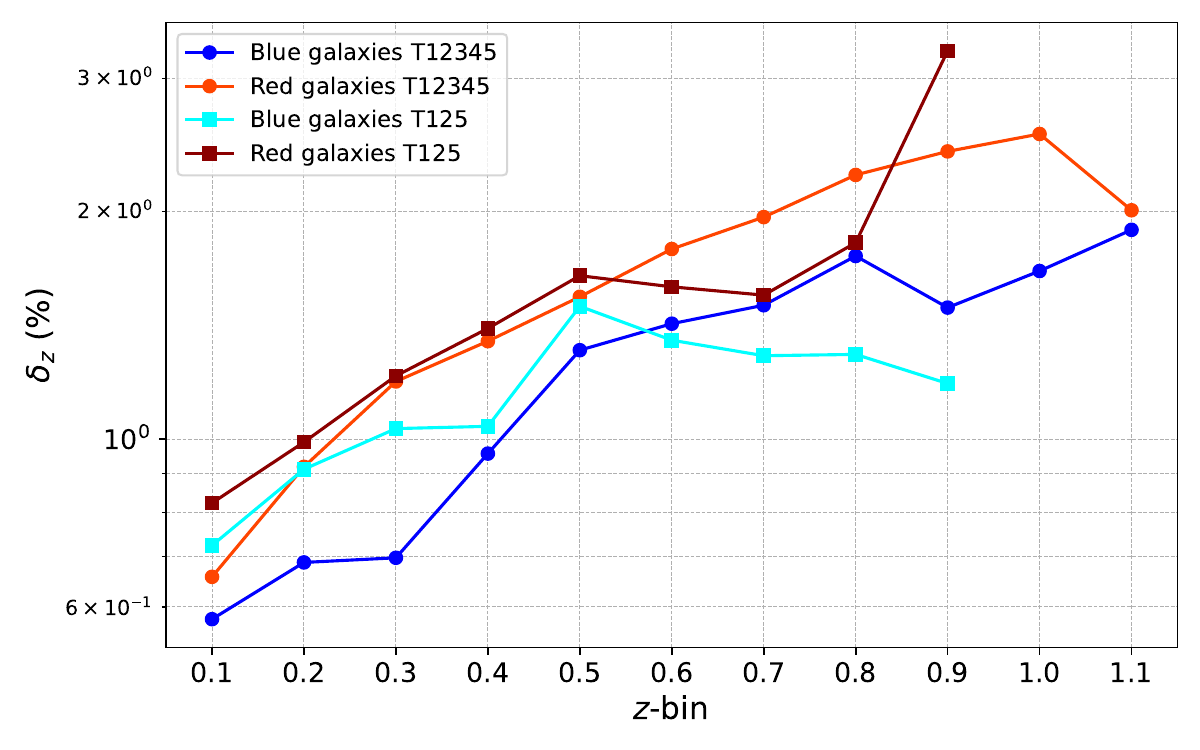}
\caption{
Galaxy number densities $n$ (left) and median photometric redshift errors $\delta_z$ (right) from the J-PAS 2024 IDR. The densities are derived with the CatBoost model and the photometric redshift errors with \texttt{LePhare}. The values are shown for different tracers and tray configurations. Note that, for the four lowest $z$-bins of red galaxies, the 3-tray configuration yields higher galaxy densities than the 5-tray setup, due to differences in the galaxy samples used to estimate them.}
\label{fig:JPAS_z_photoz}
\end{figure}

The number densities were estimated using three different approaches, all of which yielded similar results. For this work, we use the values derived with the CatBoost Model (CBM). The CatBoost algorithm \cite{CatBoostAlgorithm} consists in a gradient-boosting tree method that takes into account morphology parameters and fluxes in 54 J-PAS bands, along with 2 CatWISE bands, as well as proper motion and parallax from Gaia. This method was selected due to its good performance and for providing slightly higher density values for both J-PAS red and blue galaxies. To establish if an object's redshift belongs within a certain interval $z \pm \Delta z$ it is useful to define the photometric odds, defined as the relative area of the Probability Density Function (PDF) within that redshift interval. An odds-cut close to $1$ means that the PDF is narrowly condensed around the highest PDF value \cite{MiniJPASPhotoZ2}. The J-PAS data were divided into different odds-cut intervals, each indicating different reliabilities in object classification. We use the data from the closest to one odds-cut interval $[0.95,1]$, corresponding to the most confident object classification. The photometric redshifts for the studied objects are estimated with \texttt{LePhare} \cite{LePharePaper}. The object densities $n$ and the median photometric redshift errors $\delta_z$ are plotted in \cref{fig:JPAS_z_photoz} for each redshift bin.

The observed area of the sky $A_{\text{sky}}$ will be determined by the survey strategy, including the tray configuration, and the observation time. We assume two possible J-PAS sky coverages: a full survey case of $8500 \text{ deg}^2$ and a half survey case of $4000 \text{ deg}^2$, corresponding to sky fractions $f_{\mbox{sky}}$ of $0.206$ and $0.097$, respectively.

To extract cosmological information from LSS data it is necessary to account for galaxy bias. This complex process depends on the different tracers of the LSS, and can be derived from bottom-up or top-down approaches, considering local or non-local bias, and incorporating non-linearities. We assume a scale-independent bias model, adopting the relation in \cite{Fry1996}:
	\begin{equation}\label{biasELG}
b(z) = \frac{b_0}{D(z)}, 
	\end{equation}
with $b_0 = 0.84$ for blue galaxies \cite{JPASSpecifications} and $b_0 = 1.70$ for red galaxies \cite{BiasHighZ}, being $D(z)$ the growth factor.

Given the simplicity of this bias model, we also considered a more flexible parametrization \cite{JPASForecastNeutrinos2} with additional degrees of freedom: 
$b^{\rm red}(z) = b_0^{\rm red} (1 + c_0^{\rm red} z)/D(z)$ and \\
$b^{\rm blue}(z) = q \, b_0^{\rm red} (1 + c_0^{\rm blue} z)/D(z)$, where 
$q = b_0^{\rm blue}/b_0^{\rm red}$, and $c_0^{\rm red}$ and $c_0^{\rm blue}$ were treated as nuisance parameters and marginalized over in the analysis. 
We found that including these additional parameters produced only minor changes in the reconstructions. Consequently, we adopt the simpler bias model in \cref{biasELG} for the remainder of this work.

\section{Primordial power spectrum reconstructions}
\label{sec:PPSReconstructions}

We use a model-independent approach to reconstruct the primordial power spectrum $P_{\mathcal{R}}(k)$. This is done sampling $N$ knots freely in the log $\{k,P_{\mathcal{R}}\}$ plane, using the nested sampler PolyChord \cite{MandatoryPolyChord1,MandatoryPolyChord2}. We refer the reader to the presentation of this methodology in \cite{MethodologicalPaper} for a detailed description. A brief summary of the methodology is presented in this section.

The framework for sampling and statistical modeling, Cobaya \cite{CobayaMandatory1,CobayaMandatory2}, is used to implement the model and the likelihood of the $P_{\mathcal{R}}(k)$ reconstructions, as well as to run the sampler PolyChord. We assume that the primordial power spectrum can be represented as a series of $N$ knots linearly interpolated in logarithmic scale. The knots are pairs of coordinates\footnote{\label{FootnoteNodes} The knot coordinates that are indeed sampled are $\{x_i,y_i\}$. These coordinates are $k$-ordered and normalized according to the used scales: $x_1 = 0$ for the largest scale $k_1 = 0.02 \text{ h } \text{Mpc}^{-1}$ and $x_f = 1$ for the smallest scale $k_f = 0.2 \text{ h } \text{Mpc}^{-1}$. This is taken into account in order to solve the Label Switching Problem \cite{SwitchingLabelProblem}. The transformation into the physical coordinates $\{\text{log}(k_i), \text{log}(P_{\mathcal{R}}^{i})\}$ is straightforward. More details can be found in \cite{MethodologicalPaper}.} $\{\text{log}(k_i), \text{log}(P_{\mathcal{R}}^{i})\}$ which are sampled to obtain the posterior distribution. We consider different numbers of knots $N$, and then marginalize over this parameter. For a given $N$, the number of sampled parameters is $2N+1$: $2N-2$ knot parameters (since the outermost knot scales are not sampled) plus the three cosmological parameters $\{H_0,\Omega_b h^2,\Omega_c h^2\}$. The priors for all the sampled parameters and the fiducial cosmology are listed in \cref{tab:Priors}. PolyChord is run with $25$ live points for each sampled parameter, resulting in a total of $50N+25$ live points. In particular, the two-knot configuration for a single $z$-bin required approximately $300$ CPU hours. Adding a knot or another $z$-bin increases the computational cost by a similar amount. In total, the reconstructions presented in this work required about $2 \times 10^5$ CPU hours, using $128$ CPU cores for each sampling case.
 	\begin{table}[t]
 		\centering
 		\begin{tabular}{|c|c|}
\hline
\textbf{Sampled Parameter} & \textbf{Prior} \\
\hline
$H_0$ & Gaussian with $\sigma = 0.54\ \mathrm{km}\ \mathrm{s}^{-1}\ \mathrm{Mpc}^{-1}$ \\
$\Omega_b h^2$ & Gaussian with $\sigma = 0.00015$ \\
$\Omega_c h^2$ & Gaussian with $\sigma = 0.0012$ \\
$y_i$ & Uniform in range $[-23,-19]$ \\
$x_i$ & Uniform in range $[0,1]$ \\
\hline
\end{tabular}

\vspace{0.5cm}

\begin{tabular}{|c|c|c|}
\hline
\textbf{Fiducial Cosmology} & \textbf{Description} & \textbf{Value} \\
\hline
$H_{0,\text{fid}}$ & Hubble constant & $67.37\ \mathrm{km}\ \mathrm{s}^{-1}\ \mathrm{Mpc}^{-1}$ \\
$(\Omega_b h^2)_{\text{fid}}$ & Baryonic density & $0.02237$ \\
$(\Omega_c h^2)_{\text{fid}}$ & Cold dark matter density & $0.1200$ \\
$A_{s,\text{fid}}$ & Amplitude of $P_{\mathcal{R}}(k_0)$, with $k_0 = 0.05\ \mathrm{Mpc}^{-1}$ & $2.0905 \times 10^{-9}$ \\
$n_{s,\text{fid}}$ & Scalar spectral index & $0.9646$ \\
\hline
\end{tabular}

 		\caption{Priors used for the sampled parameters and fiducial cosmology. The cosmological parameters $\{H_0,\Omega_b h^2,\Omega_c h^2\}$ follow correlated Gaussian distributions from Planck DR3 \cite{Planck18Parameters,PlanckLegacy} centered on the fiducial cosmology parameters. The knot parameters\textsuperscript{\ref{FootnoteNodes}} $x_i$ and $y_i$ are sampled from uniform priors. We consider a spatially flat universe ($\Omega_k = 0$) and neglect the contribution of neutrinos, setting $\Omega_{\nu}$ = 0.}
			\label{tab:Priors}
	\end{table}
We reconstruct $P_{\mathcal{R}}(k)$ in the scale range $k \in [0.02,0.2] \text{ h}\text{ Mpc}^{-1}$, divided into 92 logarithmically spaced bins with $\text{log}(\Delta k) = 0.02$. The lower limit of $0.02 \text{ h}\text{ Mpc}^{-1}$ corresponds to the largest scale at which J-PAS maintains sufficient signal-to-noise ratio, while the upper limit of $0.2 \text{ h}\text{ Mpc}^{-1}$ avoids the complex non-linear regime \cite{PFinNLScales}. We adopt uniform priors for the knot parameters, and for the cosmological parameters $\{H_0,\Omega_b h^2,\Omega_c h^2\}$ we use the correlated Gaussian posteriors from Planck DR3  \cite{Planck18Parameters}, as listed in \cref{tab:Priors}. Because the feature considered here is perturbative and confined to small scales, and that the Planck constraints consider all the multipole range, thus being less sensitive to local features, the use of Standard Model-based Planck priors remains approximately valid in this context. Moreover, these priors reduce degeneracies between the knots and the cosmological parameters, thereby lowering computational costs. We also tested broader priors to assess robustness. In particular, doubling the Gaussian prior width on $H_0$, motivated by the DESI DR2 tension \cite{DESIDR2}, led to only minor changes in the reconstruction and feature recovery. Therefore, we keep the priors from Planck DR3 throughout this work.

The likelihood function $\mathcal{L}$ is computed by comparing the model galaxy power spectra $P_{g}(\mbox{model})$ to those derived from J-PAS data $P_{g}(\mbox{data})$, across different tracers, redshift bins, and the previously defined $k$-grid. The likelihood is modeled as a multivariate Gaussian:
\begin{equation}\label{likelihood}
-2 \log(\mathcal{L}) = \sum_{t} \sum_{\alpha} \sum_{\beta} \left[  \boldsymbol{x}_{t,\alpha \beta}^{T} \sigma_{t,\alpha \beta}^{-1} \boldsymbol{x}_{t,\alpha \beta} + \log \left[\det(Cov_{tt,\alpha \beta}) \right]\right],
\end{equation}
where $\boldsymbol{x}_{t,\alpha \beta} \equiv [\boldsymbol{P}^{(0)}_{g,t,\alpha \beta}(\mbox{model}) - \boldsymbol{P}^{(0)}_{g,t,\alpha \beta}(\mbox{data})]$; $t$ indexes the tracer, $\alpha$ the $z$-bins, and $\beta$ the $k$-bins. $Cov_{tt,\alpha \beta}$ is the covariance estimated from the model and includes inter-tracer correlations, but is assumed to be diagonal in both redshift and scale. According to \cite{CovMatrixAssumptions}, when the survey geometry is not known, adopting a covariance matrix diagonal in $k$ is a good approximation, even on mildly non-linear scales of $k \approx 0.2 \text{ h} \text{ Mpc}^{-1}$. Moreover, the non-Gaussian contributions identified in that work were shown to have only a marginal impact on the parameter error bars in a full-shape analysis of BOSS DR12. We therefore follow their approach and consider a Gaussian covariance matrix.

We combine the reconstructions from all $N$-knot configurations into a posterior marginalized over $N$, using evidence-dependent weights. These are obtained from the product of PolyChord’s importance weights and the normalized evidences $Z_N$ for each $N$ \cite{MethodologicalPaper}.

Our results are based on a comparison between the marginalized reconstruction and the power-law case. In order to do so, we perform two statistical tests: a Bayes factor test and a hypothesis test. The Bayes factor $Z_2/Z_{\text{max}}$ is used to evaluate whether the power-law model is preferred over more flexible reconstructions, following Jeffreys criterion \cite{Jeffreys1998theory}, where $Z_{\text{max}}$ is the maximum evidence among reconstructions with $N > 2$. For the considered features, our reconstructions are still sensitive to configurations with $Z_N/Z_{\text{max}} > 10^{-2}$. The evidences exhibit a decreasing trend as $N$ increases. Thus, exploring high-$N$ configurations is not expected to provide significant information for the reconstructions. Since configurations with $N > 7$ have evidences at least three orders of magnitude smaller than the maximum one, we consider a maximum number of nodes of $7$. We also apply a hypothesis test \cite{Cowan} comparing the marginalized and the power-law distributions at each $k$. We adopt a significance level $\alpha = 0.05$. The power of the test $1-\beta$ quantifies the separation between both distributions, allowing the classification of the feature detection status: $1-\beta > 0.9$ shows a good indication of the presence of a feature; $0.9 > 1-\beta > 0.5$ shows a hint of a possible feature; and $0.5 > 1-\beta$ shows no evidence of a feature. The Bayes factor offers a more robust global comparison, while the hypothesis test enables the localization of a feature in $k$. We quantify detection capability using the fraction of scales in which deviations occur, taking advantage of the complementary insights provided by both tests.

\section{Galaxy power spectrum model}
\label{sec:GPSModel}

In order to reconstruct the primordial power spectrum $P_{\mathcal{R}}(k)$, we simulate the observable monopole galaxy power spectrum $P^{(0)}_{g}(k,z)$ expected from J-PAS. First, we compute its mean value $\bar{P_{g}}^{(0)}(k,z)$ using a model that incorporates redshift-space distortions, non-linear damping, photometric redshift uncertainties, and geometric distortions. We detail the construction of the used galaxy power spectra model below.

The first step is to select a template for the primordial power spectrum $P_{\mathcal{R}}(k)$. We consider deviations from the standard power-law form, motivated by inflationary models that introduce Local Oscillatory (LO) features \cite{ModelAdams2001,ModelAchucarro2010} or Global Oscillatory (GO) ones \cite{ModelFlauger2017,SearchingFeatures1}. Details of these feature templates can be found in the Appendix of \cite{MethodologicalPaper}. The linear matter power spectrum $P_m(k)$ is derived from $P_{\mathcal{R}}(k)$ by means of the transfer function $T(k)$:
	\begin{equation}\label{MatterPS}
P_m(k) = 2 \pi^2 h^3 T^2(k) k P_{\mathcal{R}}(k).
	\end{equation}
This spectrum is computed using the Boltzmann code CAMB \cite{CAMB} for the different redshift bins.

To account for non-linear clustering effects that damp the Baryon Acoustic Oscillations (BAO) signal, we use a dewiggled matter power spectrum $P_{dw}(k)$, in which small-scale BAO features are smoothed. Details on the smoothing procedure and its impact on the $P_{\mathcal{R}}(k)$ reconstructions are provided in the Appendix.

Then, we compute the galaxy power spectra. We assume a Kaiser model for the redshift-space distortions \cite{RSDKaiser}, both Fingers of God \cite{FoGPaper} and Alcock-Paczynski \cite{APEffectOriginal, APEffect} effects, and a photometric error power suppression term \cite{EuclidForecastModel,PkModel}:
	\begin{equation}\label{GalaxyPS}
\bar{P_g}(k,\mu,z) = \frac{D_{A,\text{fid}}^2 E(z)} {D_A^2 E_{\text{fid}}(z)} F_{\text{FoG}}(k',\mu',z) \left[b(z)+f(z) \mu'^{2}\right]^{2} P_{dw}(k',\mu',z) e^{-k'^{2} \mu'^{2} \sigma_{z}^{2}(z)},
	\end{equation}
 with $\mu$ the cosine of the angle between the wavevector $\vec{k}$ and the line of sight. $D_A(z)$ is the angular diameter distance and $b(z)$ the previously discussed bias. For a given matter density parameter $\Omega_m$, the Hubble function is defined as $H(z) = H_0 E(z)$, with $E(z) = \sqrt{\Omega_m (1+z)^3+(1-\Omega_m)}$, and $f(z) = \left(\Omega_m (1+z)^3 \frac{1}{E^2(z)}\right)^\gamma$ the growth function, with a growth index $\gamma = 0.545$. The subscript `fid' indicates that a quantity is evaluated at the fixed fiducial cosmology (see \cref{tab:Priors}).
 
 The different effects are taken into account as follows:
 
 \begin{enumerate}

    \item The photometric redshift error $\delta_z$ suppresses the galaxy power spectrum $P_g(k)$ via the exponential factor $\text{e}^{-k^{2} \mu^{2} \sigma_{z}^{2}(z)}$, where $\sigma_{z} = \frac{\delta_z (1+z)}{H(z)}$. This suppression arises because redshift uncertainties introduce line-of-sight errors in the comoving position $r$, modeled as a Gaussian $\text{e}^{\frac{-(\Delta_r)^2}{2 \sigma_{z}^2}}$. Convolving this with the true galaxy distribution smooths density fluctuations along the line of sight, which in Fourier space translates into the exponential damping of power. The values of the median photometric errors $\delta_z$ expected for J-PAS were given in \cref{fig:JPAS_z_photoz}.

     \item The Fingers of God (FoG) effect must be included in LSS analyses, since it accounts for small-scale redshift-space distortions caused by the random motions of galaxies within virialized structures (like galaxy clusters). It is taken into account with the function $F_{\text{FoG}}(k,\mu,z)$ modeled as a Lorentzian \cite{FoGPaper}:
    \begin{equation}\label{FingersOfGod}
F_{\text{FoG}}(k,\mu,z) = \frac{1}{1+ \left[f(z) \text{ }\mu \text{ } k \text{ }\sigma_{p,\text{fid}}(z)\right]^2}, 
    \end{equation}
where the dispersion parameter $\sigma_{p,\text{fid}}(z)$ is obtained integrating the linear matter power spectrum as:
\begin{equation}\label{DispersionParameter}
\sigma^2_{p,\text{fid}}(z) =  \frac{1}{6 \pi^2} \int P_{m,\text{fid}}(k,z) \text{ d}k.
	\end{equation}
In practice, this integral is evaluated within finite boundaries, with $k_\text{min} = 10^{-5} \text{ h } \text{Mpc}^{-1}$ and $k_\text{max} = 10^{3}  \text{ h } \text{Mpc}^{-1}$.

\item The sampled cosmology will differ from the true cosmology. Observed structures (like the BAOs) appear stretched or compressed along and across the line of sight. The Alcock-Paczynski effect \cite{APEffectOriginal} corrects for this distortion by rescaling the wavenumber $k$ and $\mu$ as:
\begin{equation}\label{AlcockPaczynskiKDistortion}
\begin{aligned}
k' &= Q k, \\
\mu' &= \frac{E}{E_{\text{fid}}Q} \mu,
\end{aligned}
\end{equation}
with:
\begin{equation}\label{AlcockPaczynskiQ}
Q = \frac{\sqrt{E^2 \chi^2 \mu^2-E_{\text{fid}}^2 \chi_{\text{fid}}^2\left(\mu^2-1\right)}}{E_{\text{fid}} \chi}.
	\end{equation}
The factor $\frac{D_{A,\text{fid}}^2 E(z)} {D_A^2 E_{\text{fid}}(z)}$ must be included in the galaxy power spectrum of \cref{GalaxyPS} for modeling the Alcock-Paczynski effect properly.

 \end{enumerate}
Effects such as non-linearities or survey geometry can significantly affect galaxy power spectrum modeling. To estimate the impact of non-linearities on our reconstructions, we considered the 1-loop renormalized perturbation theory corrections \cite{RPTPaper} to the matter power spectrum, which yield a slight increase in the signal-to-noise ratio at intermediate and small scales. A realistic galaxy power spectrum modeling should include these non-linear effects, since the linear approach underestimates the power in $\approx 30\%$ at the scales of $\approx 0.1 \text{ h}\text{ Mpc}^{-1}$. Moreover, the non-linear effect affects the covariance matrix, as explained in the previous section, but to a small extent \cite{CovMatrixAssumptions}. Nevertheless, this underestimation is not biasing our results, as both the sampling model and the galaxy power spectrum data are constructed coherently. In future work, an efficient way of considering non-linearities is by adopting a parametrized model for the galaxy power spectrum that extends the validity range of our approach. Regarding survey geometry, we assume the simplest one. A more realistic survey geometry would affect non-linear scales, as it can introduce mode mixing among other effects. However, its impact on the final precision is expected to be moderate once properly accounted for. Therefore, a realistic description of non-linear scales requires both a sophisticated non-linear model and a precise description of survey geometry, which is beyond the scope of this work.

 Following \cref{GalaxyPS}, the  bias parameter $b_0$ of the bias model \cref{biasELG} is degenerate with the amplitudes of the reconstructed knots. Allowing for a varying $b_0$ would lead to the monopole galaxy power spectrum reconstructions constraining a combination of $b_0$ and the knots amplitude. Nevertheless, this degeneracy is irrelevant for feature recovery in this forecast, since we coherently quantify the prominence of the perturbative feature with respect to a power-law model. To incorporate the angular dependence encoded in $\mu$, the multipole moments of the galaxy power spectrum $P_g^{(\ell)}(k)$ can be computed by projecting it onto the Legendre polynomials, as:
\begin{equation}\label{GalaxyMultipoles}
\bar{P_{g}}^{(\ell)}(k,z)=\frac{(2 \ell+1)}{2} \int_{-1}^{1} \bar{P_{g}}\left(k, \mu, z \right) \mathcal{L}_{\ell}(\mu) d \mu,
	\end{equation}
with $\mathcal{L}_{\ell}(\mu)$ the Legendre polynomial of degree $\ell$. The monopole ($\ell=0$) is: 
\begin{equation}\label{GalaxyMonopole}
\bar{P_{g}}^{(0)}(k,z)=\frac{1}{2} \int_{-1}^{1} \bar{P_{g}}\left(k, \mu, z \right) d \mu.
\end{equation}

Nevertheless, we perform the $P_{\mathcal{R}}(k)$ reconstructions considering only the monopole galaxy power spectrum, since the contribution from higher-order multipoles is expected to be small for the isotropic signals generated by inflation. The small leakage expected from the application of a window function, together with the intrinsically smaller signal-to-noise of the higher-order multipoles with respect to the monopole, makes the expected net effect on the feature recovery small. In addition, the feature amplitude is expected to be small, which makes the possible contribution from higher-order multipoles less relevant.

To estimate the $\bar{P_{g}}^{(0)}$ uncertainties, we compute its covariance matrix. We follow \cite{CovarianceMatrix}, which accounts for the Fourier number of modes assigned to the $k$-shell, $N_k(z)$. In order to compute this number, we define the comoving radial distance as:
	\begin{equation}\label{ComovingDistance}
\chi(z) = \frac{1}{H_0} \int_0^z\frac{1}{E(z')}d z',
	\end{equation}
so the volume of a redshift bin of mean value $z$ with upper and lower bounds $z_+$ and $z_-$ is:
	\begin{equation}\label{VolumeAlpha}
V_\alpha(z) = \frac{4}{3}\pi f_{\mbox{sky}} \left[\chi^3(z_+)-\chi^3(z_-)\right].
	\end{equation}
Inside a $k$-shell with upper and lower scales $k_+$ and $k_-$, the number of Fourier modes $N_k$ for a redshift bin $z$ are:
	\begin{equation}\label{NumberOfModes}
N_k(z) = \frac{V_\alpha(z)}{(2\pi)^3} \frac{4}{3}\pi(k_+^3-k_-^3).
	\end{equation}
This number of Fourier modes contributes to the cosmic variance. The shot noise is computed as the inverse of the densities $n$ for each object, which were given in \cref{fig:JPAS_z_photoz}. Finally, the covariance of the monopole accounting for both effects is \cite{RescoMaroto,TesisResco}:
	\begin{equation}\label{CovMatDefinition}
Cov(k,z) =  \frac{1}{N_k(z)}\int_{-1}^{1}\left(\bar{P_{g}}(k,z,\mu)+\frac{1}{n}\right)^2  \text{d} \mu.
	\end{equation}
We can compute both the auto- and cross-correlation galaxy power spectra of different tracers $t$ and $t'$. \Cref{GalaxyPS} can be generalized as:
\begin{equation}\label{GalaxyPSTracer}
\begin{aligned}
\bar{P}_{g,tt'}(k,\mu,z) = \frac{D_{A,\text{fid}}^2 E(z)} {D_A^2 E_{\text{fid}}(z)} F_{\text{FoG}}(k',\mu',z) \times \\
\times \left[b_t(z)+f(z) \mu'^{2}\right] \left[b_{t'}(z)+f(z) \mu'^{2}\right] P_{m}(k',z) e^{\frac{-k'^{2} \mu'^{2} \sigma_{z,t}^{2}(z)}{2} } e^{\frac{-k'^{2} \mu'^{2} \sigma_{z,t'}^{2}(z)}{2} },
\end{aligned}
\end{equation}
and the covariance of \cref{CovMatDefinition} is also generalized following:
\begin{equation}\label{CovMatDefinitionTracers}
 Cov_{tt'}(k,z) =  \frac{1}{N_k(z)}\int_{-1}^{1}\left(\bar{P}_{g,tt'}(k,z,\mu)+\frac{\delta_{tt'}}{\sqrt{n_t n_{t'}}}\right)^2  \text{d} \mu,
	\end{equation}
where $\delta_{tt'}$ is the Kronecker delta.

At the largest scales the cosmic variance tends to be the dominant source of uncertainty, whereas the shot noise mainly affects the small scales. The sensitivity of the reconstructions depends on the signal-to-noise ratio, whose scale dependence across all redshift bins is discussed in the next section.

\section{J-PAS galaxy power spectrum sensitivity}
\label{sec:SNSensitivity}

\begin{figure}[t]
\centering 
\includegraphics[width=.98\textwidth]{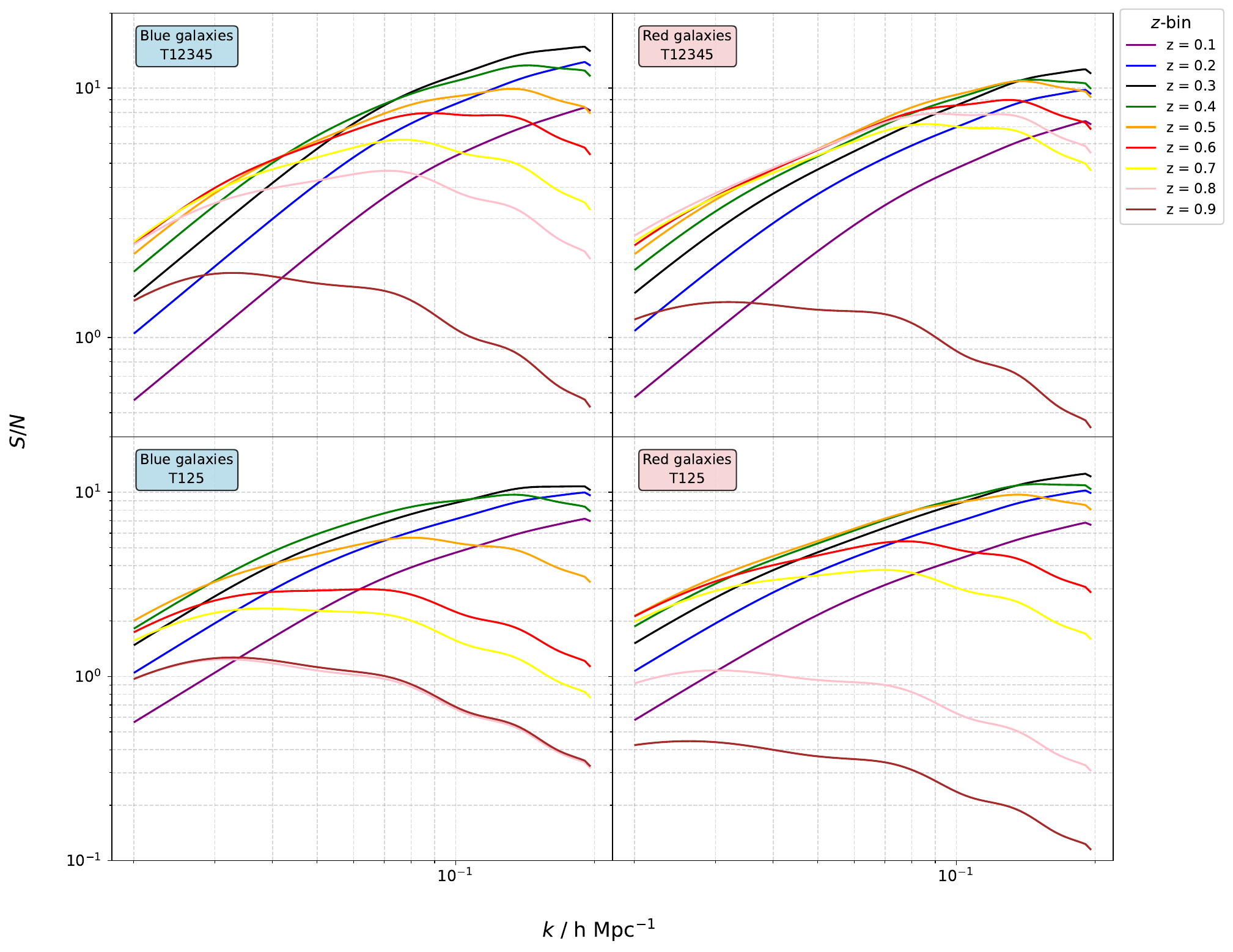}
\caption{
Signal-to-noise ratios $S/N$ of the J-PAS galaxy power spectra across redshift bins. Left panels show results for blue galaxies; right panels for red galaxies.  
Top panels correspond to the T12345 tray strategy; bottom panels to the T125 strategy. All results assume a primordial power-law model and a survey area of $8500\,\mathrm{deg}^2$.
}
	 \label{fig:SNMonopole}
\end{figure}

The sensitivity to primordial power spectrum deviations from the standard power-law depends on the signal-to-noise ratio $S/N$ of the galaxy power spectrum. This $S/N$ is illustrated in \cref{fig:SNMonopole} for the different redshift bins, tracers and tray strategies of J-PAS. As shown in the figure, the peak of the $S/N$ shifts towards larger scales as the $z$-bin increases. This peak reaches $S/N \approx 10$ at scales of approximately $0.2 \text{ h }\mathrm{ Mpc}^{-1}$ for intermediate redshift bins, and decreases by an order of magnitude at the largest scales. This trend is observed across both tray strategies and tracers. According to \cite{FAROPaper}, the relative errors in the determination of $P_g^{(0)}(k)$ in J-PAS are smaller than those of DESI and Vera C. Rubin Observatory Legacy Survey of Space and Time  (LSST) at $k \approx 0.1 \text{ h} \text{ Mpc}^{-1}$, and comparable to Euclid. Some of these missions will explore larger areas, thus reducing the sampling variance at large scales. In order to optimize the precision of the reconstruction from J-PAS, the simulated feature was placed at $k \approx 0.1 \text{ h }\mathrm{ Mpc}^{-1}$. The T12345 tray strategy yields slightly higher $S/N$ values than the T125. While red galaxies generally exhibit lower $S/N$ than blue galaxies under the T12345 strategy, a slight improvement over blue galaxies is observed for the T125 strategy. As a reference for single z-bin reconstructions, we consider the redshift bin $z = 0.4$, since it offers one of the best $S/N$ performances across the full range of scales. 

\begin{figure}[t]
\centering 
\includegraphics[width=.49\textwidth]{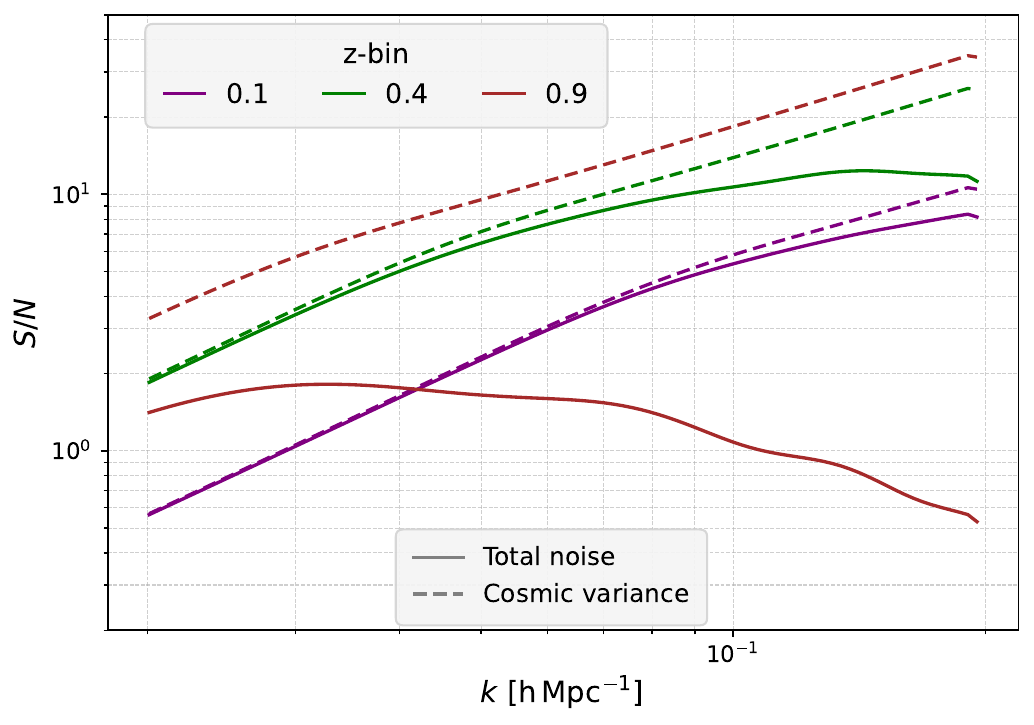}
\includegraphics[width=.49\textwidth]{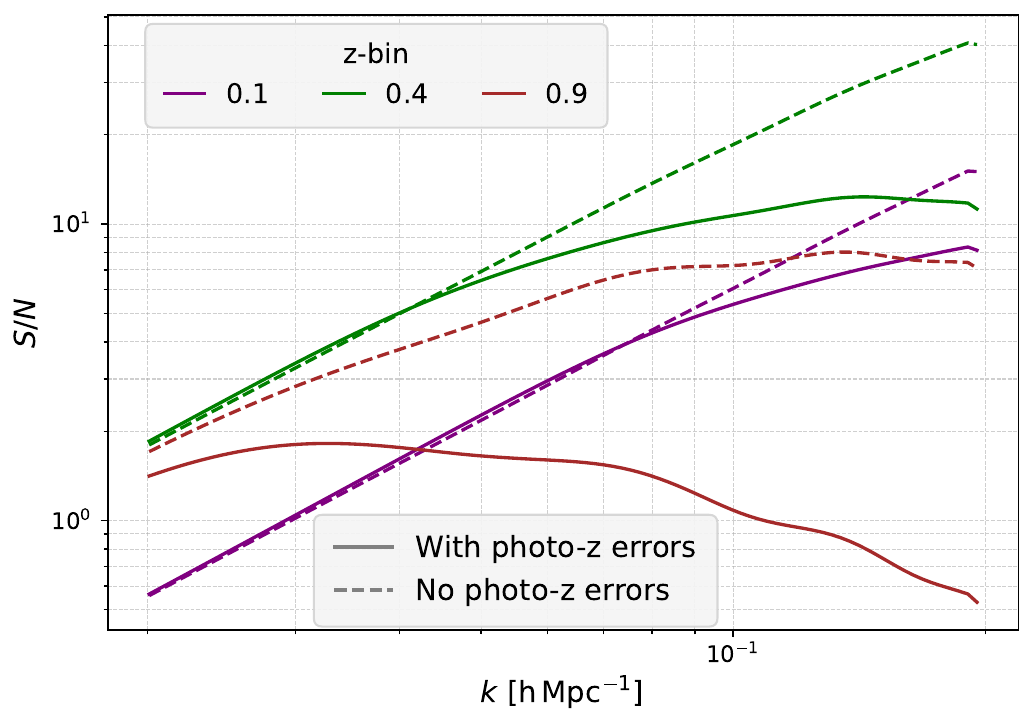}
\includegraphics[width=.49\textwidth]{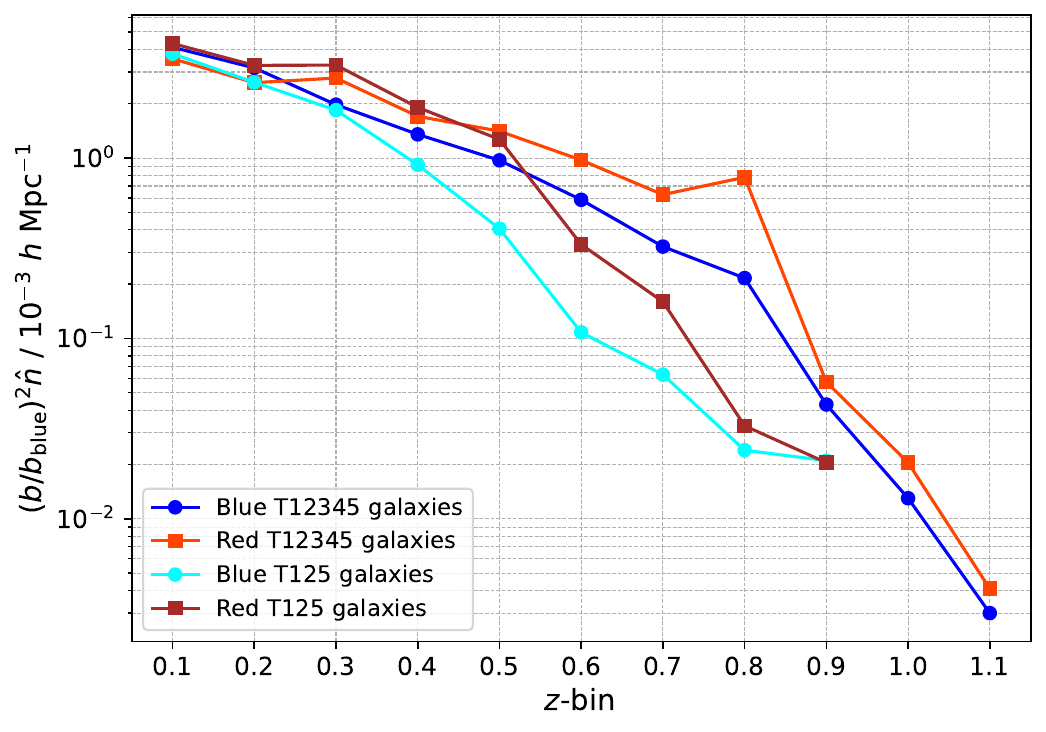}
\includegraphics[width=.50\textwidth]{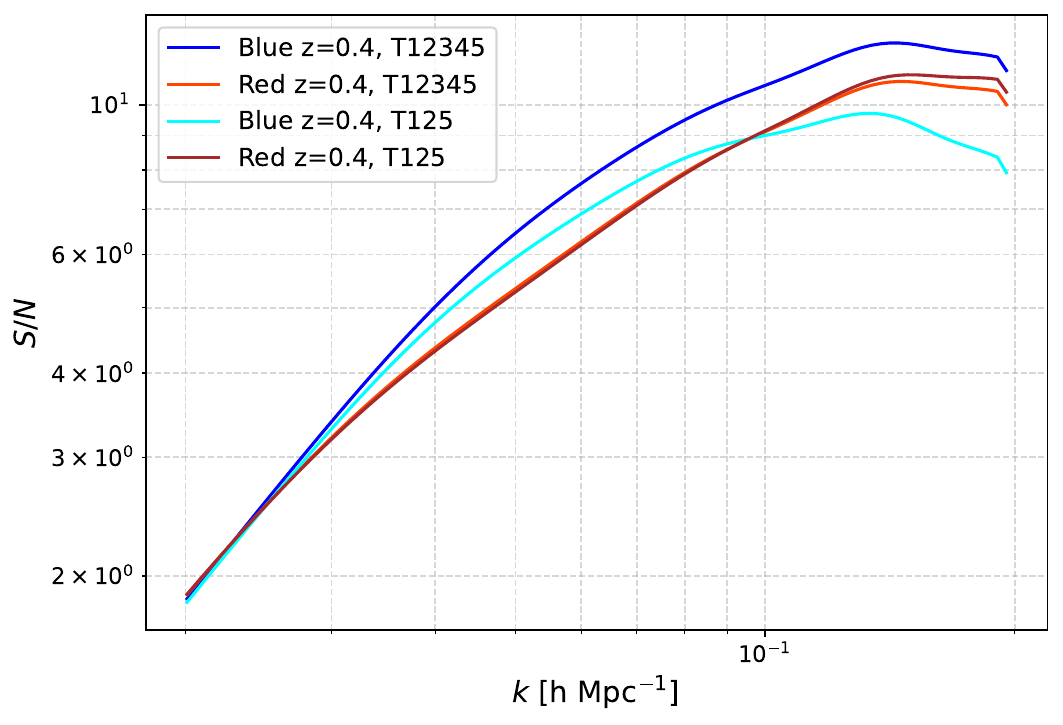}
\caption{
Top left panel: signal-to-noise ratio $S/N$ for the cosmic variance component (dashed) and the total covariance (solid). 
Top right panel: comparison of $S/N$ with (solid) and without (dashed) photometric redshift errors. We plot three representative redshift bins corresponding to low ($z = 0.1$), intermediate ($z = 0.4$), and high ($z = 0.9$) redshift. These top panels correspond to blue galaxies with $8500 \deg^2$ under the T12345 strategy. Bottom left panel: effective number density of red galaxies relative to the blue ones. Bottom right panel: $S/N$ comparison of different tracers and tray strategies at the $z = 0.4$ bin.}
\label{fig:SNAnalysis}
\end{figure}

In \cref{fig:SNAnalysis}, we analyze the different covariance components contributing to the $S/N$, and assess the impact of photometric redshift errors, bias, number densities, and tray strategies.
In the top-left panel of \cref{fig:SNAnalysis}, we show the signal-to-noise ratio evaluated under two covariance matrix assumptions: one considering only cosmic variance, and the other including the full covariance as defined in  \cref{CovMatDefinition}. From this comparison, it can be seen that J-PAS approaches the cosmic-variance limit at the largest scales in the lowest $z$-bins. At low redshifts, cosmic variance dominates, while shot noise contributes significantly only at the smallest scales. For intermediate redshifts, both components of the covariance matrix play a significant role, whereas at high redshift shot noise dominates across all scales. This increasing impact of shot noise with redshift is primarily due to the decreasing galaxy number density, while the larger volume of higher redshift bins helps mitigate cosmic variance. As a result, an intermediate bin such as $z = 0.4$ offers a good balance between these two sources of uncertainty, motivating its use in the single-bin reconstructions. Comparable $S/N$ values are found throughout the range $0.3 < z < 0.6$, a redshift interval that has been identified as the most suitable for testing modified gravity and dark energy with J-PAS \cite{JPASForecastDarKEnergyModifyGravity}.

The impact of photometric redshift errors is shown in the top right panel of \cref{fig:SNAnalysis}. It follows a similar trend as the shot noise: increasing with redshift and primarily affecting the smallest scales. At these scales, the signal-to-noise ratio is reduced by approximately 40\% in the low redshift bin, 75\% for the intermediate bin, and an order of magnitude in the high redshift bin. Although the suppression of $S/N$ due to photometric errors is more pronounced at intermediate redshifts than at low ones, it does not fully compensate for the higher cosmic variance present at low redshift.

The dependence of the $S/N$ on galaxy number density can be captured by an effective number density, defined as $b^2 n$. Although red galaxies have lower number densities than blue galaxies (as shown in the left panel of \cref{fig:JPAS_z_photoz}), their bias is approximately twice as large. As a result, the effective number density for red galaxies is enhanced by a factor of four. This quantity is shown in the bottom left panel of \cref{fig:SNAnalysis} normalized to blue galaxies in order to compare both tracers. The effective number densities of blue and red galaxies are relatively similar in the low redshift range. Consequently, at those bins photometric redshift errors become the dominant factor in determining which tracer yields a higher $S/N$. As shown in the right panel of \cref{fig:JPAS_z_photoz}, red galaxies suffer from photometric redshift errors roughly twice as large as those of blue galaxies. This typically results in blue galaxies achieving higher $S/N$ values.

The bottom right panel of \cref{fig:SNAnalysis} presents the signal-to-noise ratio for the $z = 0.4$ bin, comparing different tracers and tray strategies. Blue galaxies observed under the T12345 strategy yield the highest $S/N$ across all relevant scales. Under the T125 strategy, they also outperform red galaxies up to $k = 0.095 \text{ h } \mathrm{Mpc}^{-1}$, primarily due to their photometric redshift uncertainties, despite having lower effective number densities. For red galaxies, the differences between tray strategies are less pronounced: both strategies result in similar densities and photometric redshift errors, leading to similar $S/N$ performance. Blue galaxies under the 5-tray strategy show more sensitivity to photometric redshift precision and sky area, particularly at the scales where the feature is most prominent. In contrast, sensitivity to number density is only significant at $k \approx 0.2 \text{ h } \mathrm{Mpc}^{-1}$, making its overall impact on $P_{\mathcal{R}}(k)$ reconstructions secondary.

While this analysis focuses on the $z = 0.4$ bin, the conclusions do not directly generalize to other redshifts. As redshift increases, variations in number density and photometric redshift errors have a more pronounced effect on the signal-to-noise ratio, as illustrated in the top panels of \cref{fig:SNAnalysis}. Accordingly, the $S/N$ differences between tracers and tray strategies also become more significant with redshift: below $z = 0.3$, differences remain below 50\%; at $z = 0.4$, they can reach a factor of 2; and at the highest redshifts, they grow to nearly an order of magnitude. Across all $z$-bins, the 5-tray strategy T12345 consistently provides higher $S/N$ for both tracers, with the most notable improvement seen for blue galaxies. This trend is already evident at $z = 0.4$ bin, although red galaxies perform similarly.

\section{Results}
\label{sec:Results}

In this section, we present the results of the primordial power spectrum reconstruction based on the statistical tests. We begin by comparing them for different J-PAS observational specifications using a single redshift bin $z = 0.4$, as motivated in the previous section. We then explore the minimum amplitude of oscillatory features that can be detected by combining multiple redshift bins and tracers.

\subsection{Reconstructions with different observational specifications}
\label{subsec:ResultsSpecifications}

\begin{figure}[t]
\centering 
\includegraphics[width=.70\textwidth]{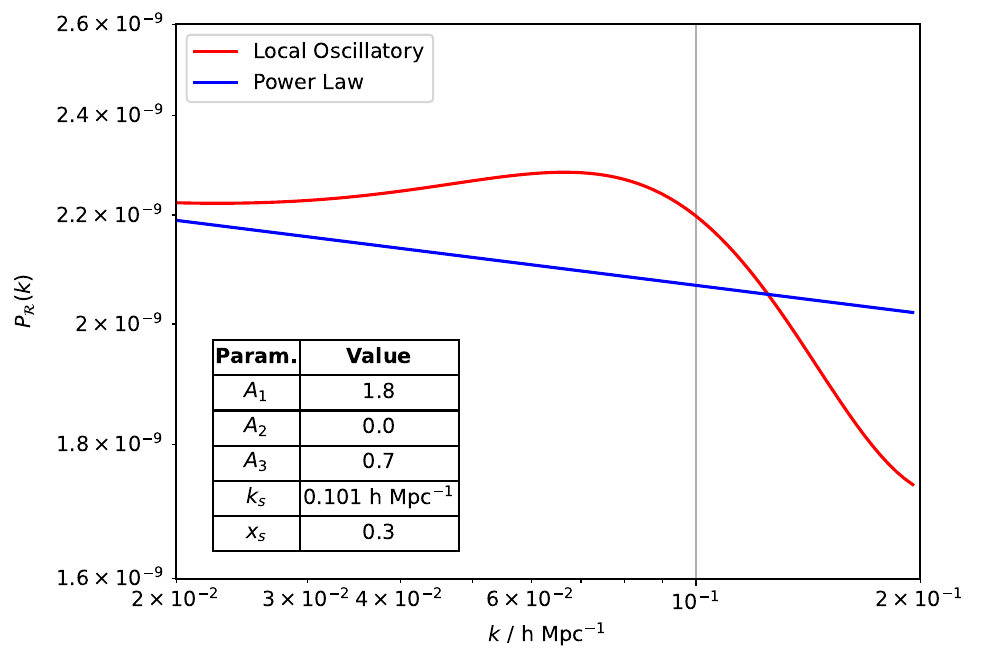}
\caption{Local oscillatory feature template (red), exhibiting an oscillation of 10\% amplitude w.r.t. the power law (blue). The values of the feature parameters are given in the table inside the plot (see appendix of \cite{MethodologicalPaper}).}
	 \label{fig:LOJPASConfigurationPlot}
\end{figure}

\begin{figure}[t]
  \centering
  \includegraphics[width=0.8\textwidth, height=8cm]{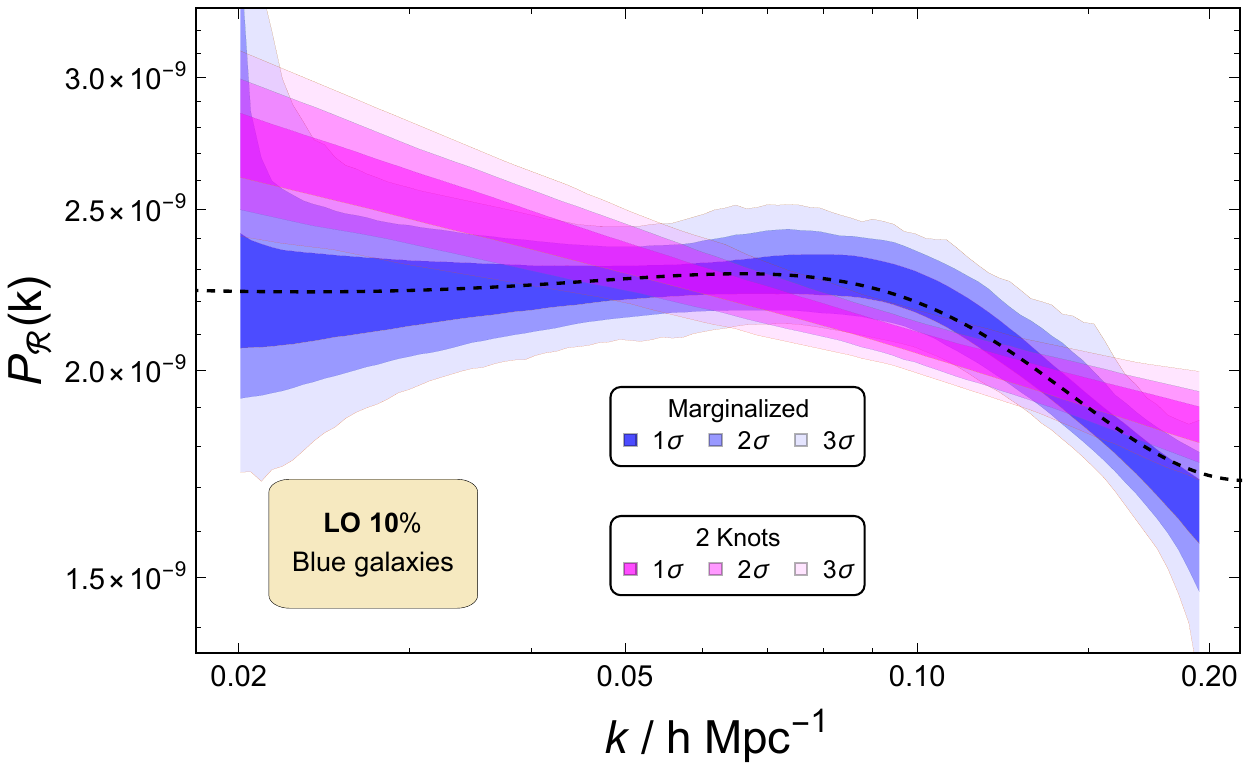}
  
\vspace{0.2cm}
   
  \includegraphics[width=0.49\textwidth, height=4.6cm]{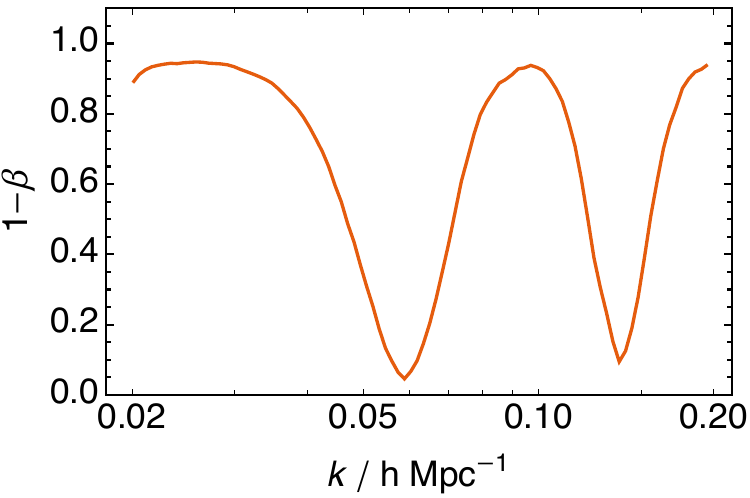}
  \hfill
  \includegraphics[width=0.49\textwidth, height=4.5cm]{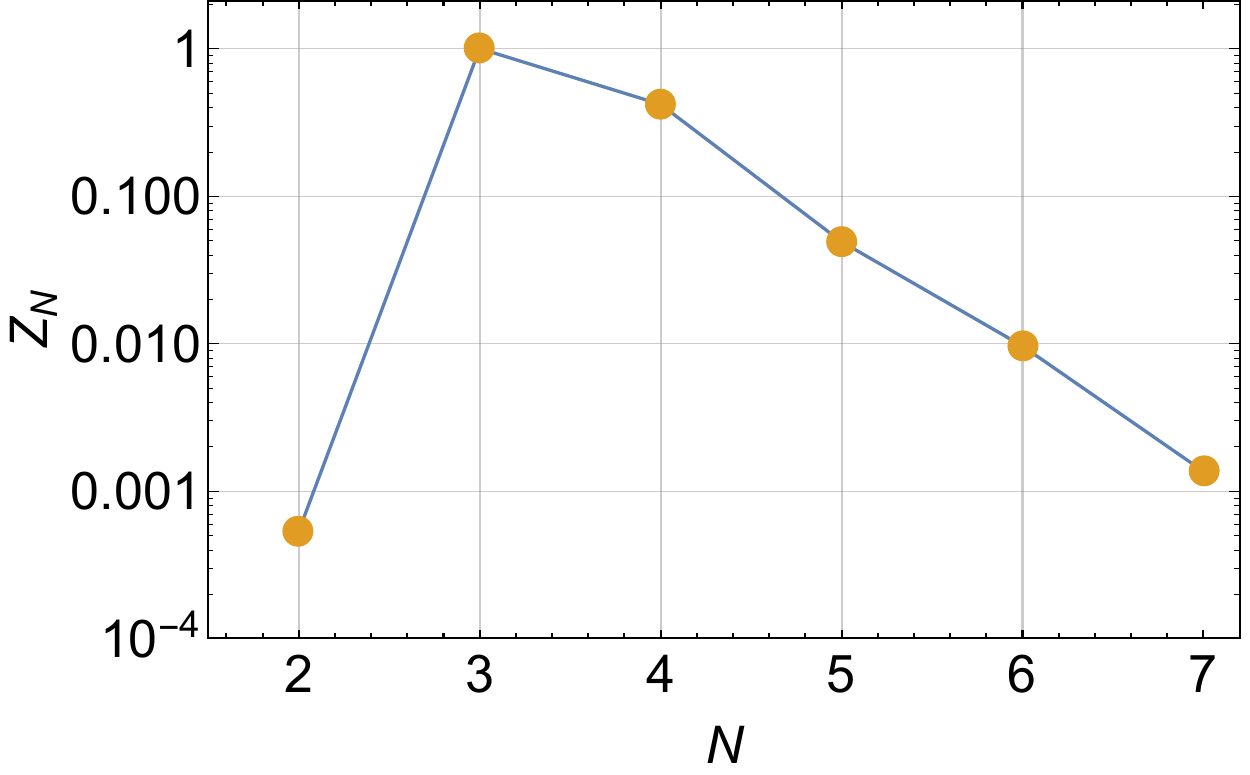}

  \caption{Top: contours of the reconstructed primordial power spectrum $P_{\mathcal{R}}(k)$ for blue galaxies at the $z = 0.4$ bin, using the T12345 strategy over $8500\,\text{deg}^2$. The input feature is a 10\% local oscillatory feature (LO 10\%, black dashed line). The magenta contours show the $N = 2$ power-law reconstruction; the blue contours represent the reconstruction marginalized over $N$. Bottom left: power of the hypothesis test. 72\% of scales exceed the 0.5 ``hint of detection" threshold, and 32\% exceed the 0.9 ``detection" threshold. Bottom right: corresponding evidences $Z_N$ as a function of the number of knots $N$. The evidences are normalized to have their maximum value equal to 1.}
    \label{fig:JPAS_Blue_DeWiggle_LO_T12345_8500_CBM_Highest}
\end{figure}

We explore how a local oscillatory feature with 10\% amplitude relative to the power-law spectrum (\cref{fig:LOJPASConfigurationPlot}) is recovered for a single $z$-bin under different J-PAS configurations. These include galaxy tracer type, survey area, and tray strategy. We select the $z = 0.4$ bin as a representative intermediate $z$-bin, as discussed in the previous section. The results of the reconstruction tests are summarized in \cref{tab:JPAS_SingleBinResults}, with the most precise $P_{\mathcal{R}}(k)$ reconstruction plotted in \cref{fig:JPAS_Blue_DeWiggle_LO_T12345_8500_CBM_Highest}. As can be seen in this figure, the Bayes factor $\frac{Z_{2}}{Z_{3}}$ indicates evidence accounting for the input feature. From the results in \cref{tab:JPAS_SingleBinResults}, we identify three primary factors influencing reconstruction performance: survey area, tracer, and tray strategy.
\begin{table}[t]
    \centering
    \begin{tabular}{|c|c|c|c|c|}
\hline
\multicolumn{5}{|c|}{\textbf{10\% Local Oscillatory feature}: single $z$-bin reconstructions} \\
\hline
\textbf{Tracer} & \textbf{Area}/$\text{deg}^2$ & \textbf{Tray} & \textbf{Bayes factor} & \textbf{Hypothesis test} \\
\hline
\multirow{4}{*}{Blue} & \multirow{2}{*}{$8500$} & T12345 & $5 \times 10^{-4}$ (Decisive) & $32$\% \& $72$\% \\
 &  & T125   & $5 \times 10^{-3}$ (Decisive) & $4$\% \& $66$\% \\
\cline{2-5}
 & \multirow{2}{*}{$4000$} & T12345 & $1.7 \times 10^{-2}$ (Very Strong) & $0$\% \& $63$\% \\
 &  & T125 & $4.8 \times 10^{-2}$ (Strong) & $0$\% \& $39$\% \\
\hline
\multirow{4}{*}{Red} & \multirow{2}{*}{$8500$} & T12345 & $1.3 \times 10^{-3}$ (Decisive) & $7$\% \& $66$\% \\
 &  & T125   & $1.5 \times 10^{-3}$ (Decisive) & $8$\% \& $65$\% \\
\cline{2-5}
 & \multirow{2}{*}{$4000$} & T12345 & $8.1 \times 10^{-2}$ (Strong) & $0$\% \& $45$\% \\
 &  & T125   & $3.1 \times 10^{-2}$ (Very Strong) & $0$\% \& $52$\% \\
\hline
\end{tabular}

\caption{
Bayes factor and hypothesis test results for the $P_{\mathcal{R}}(k)$ reconstructions of J-PAS galaxies featuring a 10\% local oscillatory deviation. Results are shown for different tracers (blue or red galaxies), survey areas ($8500 \text{ deg}^2$ or $4000 \text{ deg}^2$), and tray strategies (T125 or T12345). All analyses use the $z = 0.4$ bin and the CBM classification algorithm in the highest odds-cut. The hypothesis test reports the fraction of scales exceeding the ``detection" threshold ($1 - \beta > 0.9$) and the ``hint of detection" threshold ($1 - \beta > 0.5$).
}
\label{tab:JPAS_SingleBinResults}
\end{table}

The 10\% feature is detected with varying levels of statistical evidence. In the large-area configuration, the detection is decisive despite the tray strategy, while in the small-area case it ranges from strong to very strong, highlighting the importance of survey area. This is expected since the covariance scales inversely with the survey volume.

Comparing tracers, blue galaxies under the T12345 tray strategy yield narrower reconstructions than red galaxies, consistent with the signal-to-noise analysis presented in the previous section (bottom right panel of \cref{fig:SNAnalysis}). For the T125 strategy, red galaxies achieve stronger Bayes factors, indicating that the reconstruction is most sensitive at the scale where the feature peaks, $k \approx 0.15$.

The impact of tray strategy depends on the tracer. For blue galaxies, the Bayes factor improves by an order of magnitude in the large-area case and by nearly a factor of three in the small-area case when switching from T125 to T12345. In contrast, for red galaxies both strategies perform similarly in the large-area case, but in the small-area case, the 3-tray strategy T125 is slightly favored---again consistent with \cref{fig:SNAnalysis}.

Repeating the analysis for a feature with 5\% amplitude, we find no significant detection for any observational configuration, neither in terms of the Bayes factor, nor in the hypothesis test. Detecting such small features requires combining information from multiple tracers and redshift bins, as discussed in the next subsection.

\subsection{Minimum detectable feature amplitude}
\label{subsec:ResultsMinimum}

\begin{figure}[t]
\centering 
\includegraphics[width=.70\textwidth]{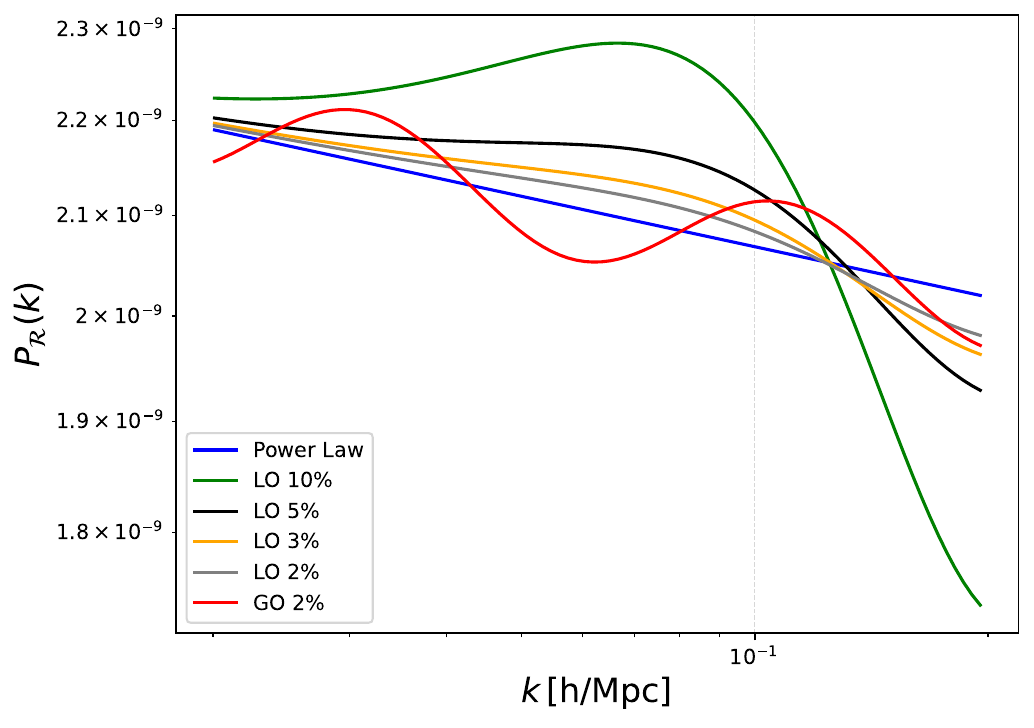}
\caption{Local Oscillatory (LO) feature template with an amplitude of oscillation of the 10\%, 5\%, 3\% and 2\% relative to the primordial power-law spectrum (in blue). A Global Oscillatory (GO) feature with 2\% amplitude is represented in red. The values of the pair of parameters $\{\mathcal{A}_1,\mathcal{A}_3\}$ used to generate those amplitudes were 
$\{1.8,0.7\}, \{0.75,0.167\}, \{0.45,0.1\}$ and $\{0.3,0.067\}$, respectively. The amplitude $\mathcal{A}_2 = 0$ in all cases, and the other parameters of the feature template are fixed at $k_s = 0.101 \text{ } \text{h} \text{ Mpc}^{-1}$ and $x_s$ = 0.3. For the GO feature, we used $A_{\text{log}} = 0.02$, $\omega_{\text{log}} = 5$, and $\phi_{\text{log}} = 0$ (see appendix of \cite{MethodologicalPaper} for details).}
	 \label{fig:LOAmplitudeFeature}
\end{figure}

We now explore the minimum amplitude of the oscillatory feature detectable with J-PAS galaxies. Specifically, we consider amplitudes of 5\%, 3\%, 2\%, and 1\% relative to the power-law spectrum, as illustrated in \cref{fig:LOAmplitudeFeature}. To maximize sensitivity, we combine redshift bins from $z = 0.1$ to $z = 0.8$ using galaxies selected from the CBM classification algorithm with the highest odds-cut. The analysis adopts the T12345 strategy over the full $8500 \text{ deg}^2$ survey area. The bins from 0.9 to 1.1 are excluded due to their low $S/N$ and to reduce computational cost. All redshift bins are assumed to be statistically independent.
To further enhance detectability, we combine both blue and red galaxy tracers. Their correlations are included following \cref{GalaxyPSTracer,CovMatDefinitionTracers}. Jointly sampling the redshift bins for both tracers yields the tightest constraint on the minimum detectable amplitude.

Results are summarized in \cref{tab:CombinedBinsResults}. At $z = 0.4$, combining both tracers is insufficient to detect the $5\%$ feature, although its Bayes factor is closer to the detection threshold than in the single-tracer cases.

\begin{figure}[t]
  \centering
  \includegraphics[width=0.8\textwidth, height=8cm]{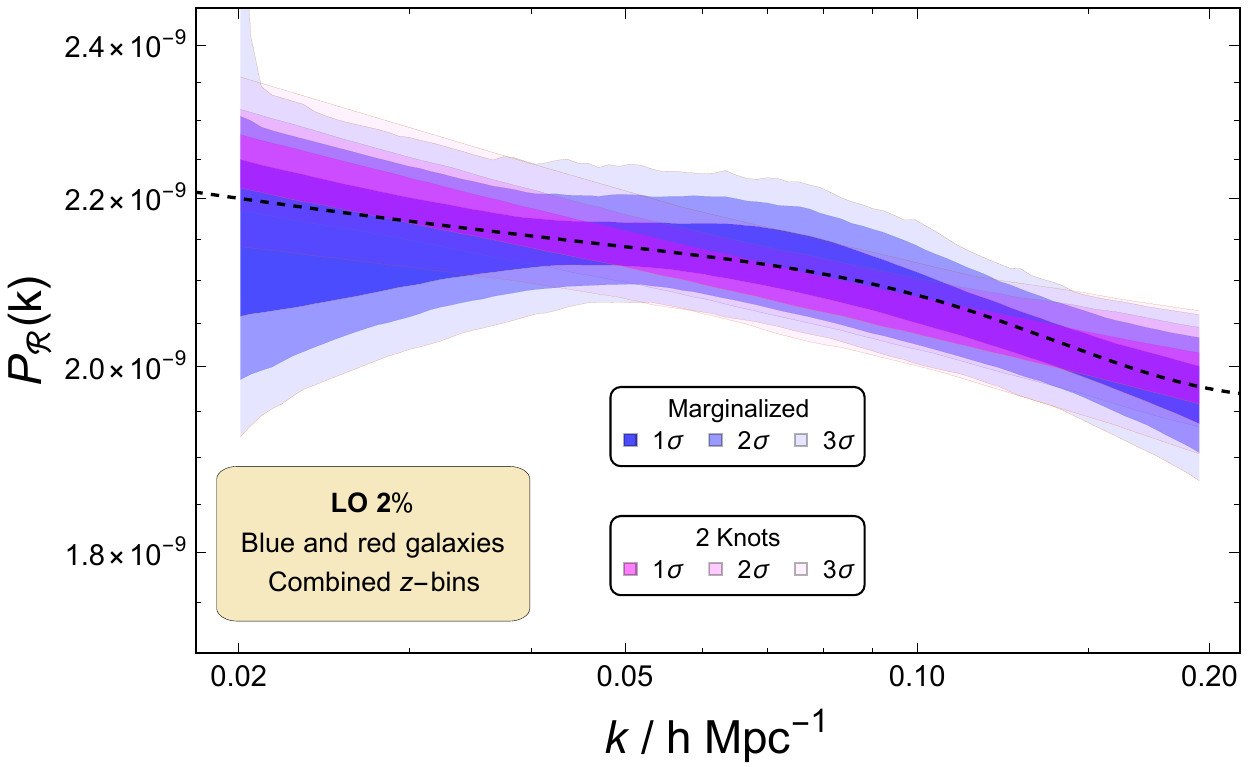}
  
\vspace{0.2cm}
   
  \includegraphics[width=0.49\textwidth, height=4.6cm]{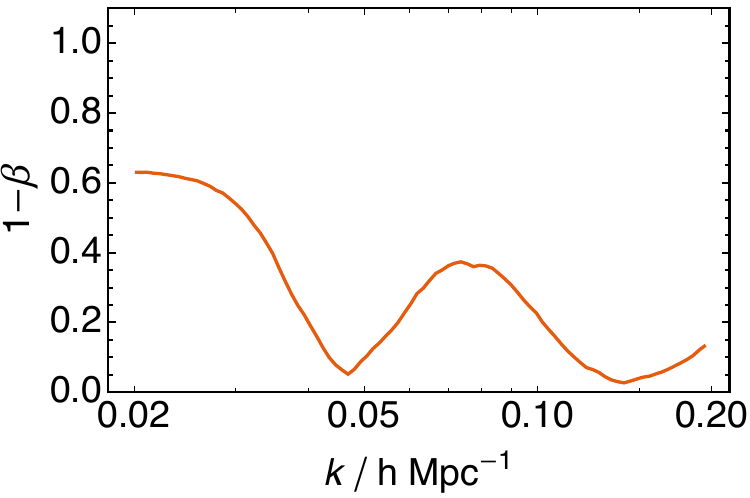}
  \hfill
  \includegraphics[width=0.49\textwidth, height=4.5cm]{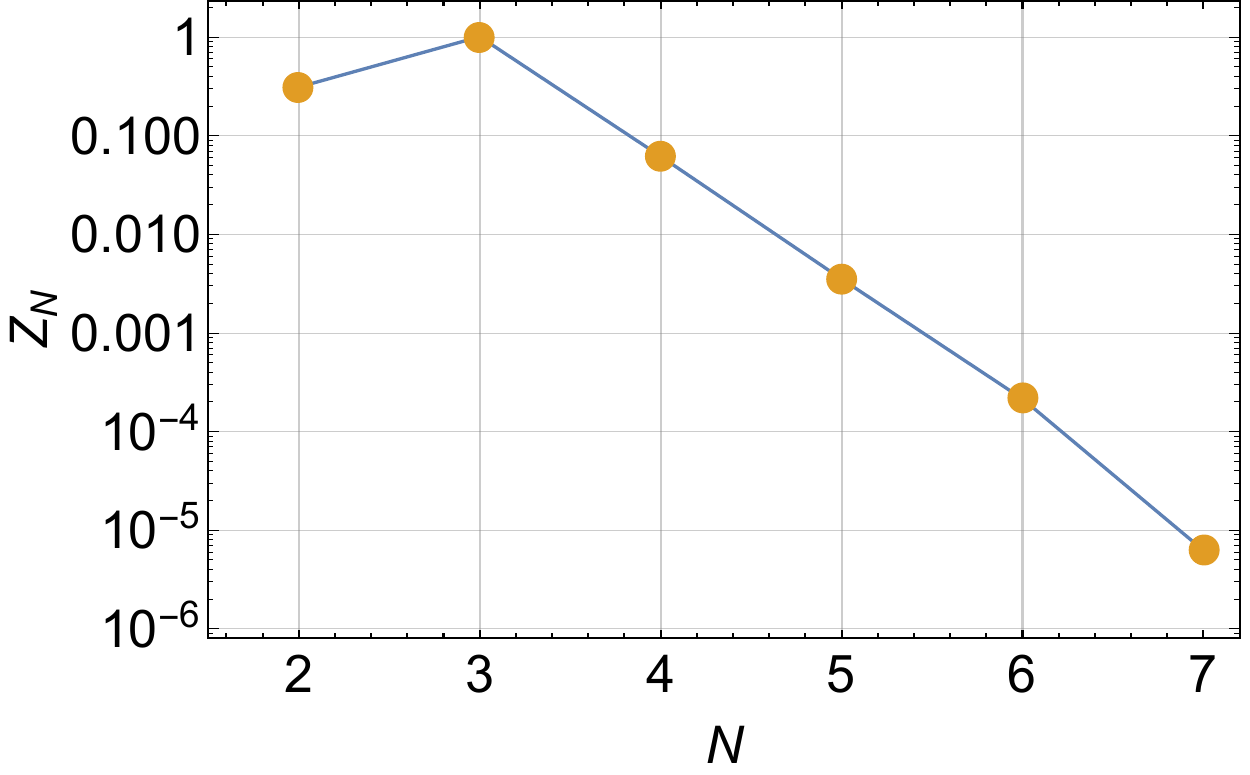}

  \caption{Top: contours of the reconstructed primordial power spectrum $P_{\mathcal{R}}(k)$ combining blue and red galaxy redshift bins, using the T12345 strategy over $8500\,\text{deg}^2$. The black dashed line shows the injected 2\% local oscillatory feature (LO 2\%)---the smallest amplitude feature recovered in this work. Magenta contours show the $N = 2$ power-law reconstruction, while blue contours represent the reconstruction marginalized over $N$. Bottom left: The hypothesis test shows that 21\% of scales exceed the 0.5 ``hint of detection" threshold, and none exceed the 0.9 ``detection" threshold. Bottom right: Bayesian evidences $Z_N$ for the same case, normalized such that their maximum value is 1. The Bayes factor $Z_2/Z_3$ shows substantial  evidence in favor of the model with the injected feature.}
    \label{fig:JPAS_CombinedTracers_AllBins_DeWiggle_LO2_T12345_8500_CBM_Highest}
\end{figure}

\begin{figure}[h]
  \centering
  \includegraphics[width=0.8\textwidth, height=8cm]{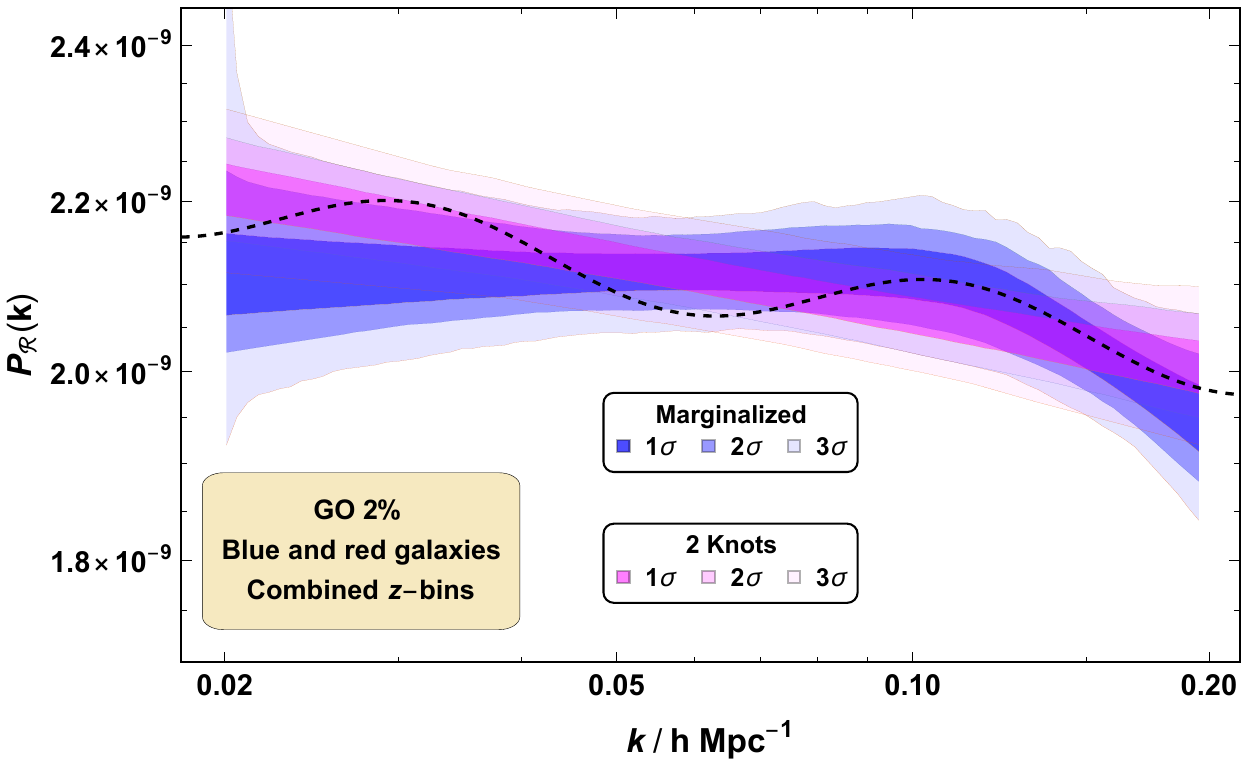}
\vspace{0.2cm}
  \includegraphics[width=0.49\textwidth, height=4.6cm]{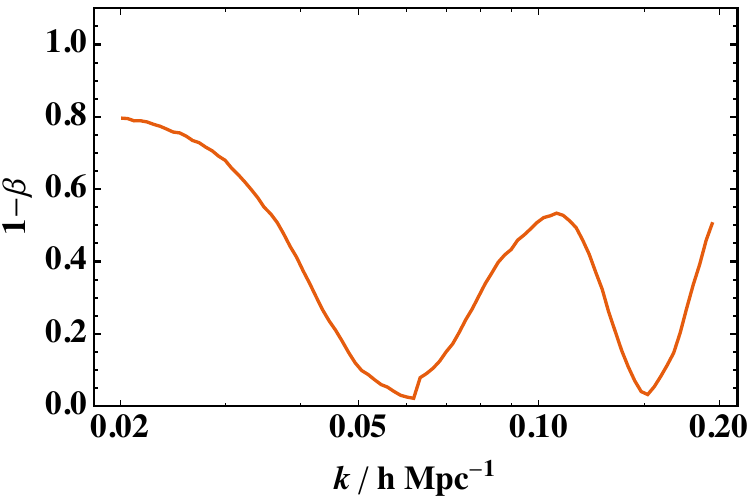}
  \hfill
  \includegraphics[width=0.49\textwidth, height=4.5cm]{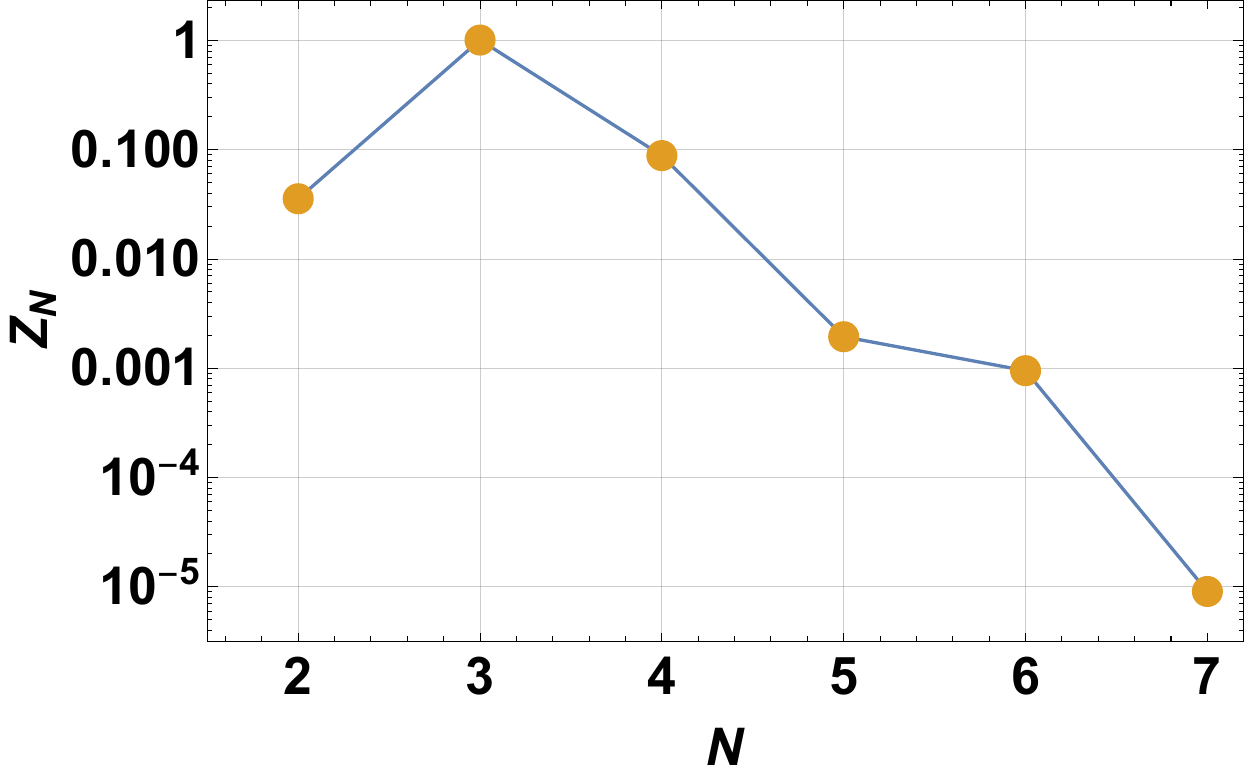}
  \caption{$P_{\mathcal{R}}(k)$ reconstruction for an input feature of a 2\% global oscillatory feature (GO 2\%, black dashed line). The panels are distributed as in \cref{fig:JPAS_CombinedTracers_AllBins_DeWiggle_LO2_T12345_8500_CBM_Highest}. The hypothesis test indicated that 34\% of scales exceed the 0.5 ``hint of detection" threshold, and 0\% exceed the 0.9 ``detection" threshold.  A substantial evidence for the feature is obtained in the Bayes factor, as in the 2\% local oscillatory feature.}
    \label{fig:GlobalFeature}
\end{figure}

When combining redshift bins separately for blue and red galaxies, the 5\% feature is decisively detected, with stronger evidence for blue galaxies. When redshift bins for both tracers are combined, a feature with $3\%$ amplitude is detected with very strong statistical significance, and a 2\% feature is recovered with substantial evidence for both local and global features. The reconstruction for the 2\% local oscillatory feature is shown in \cref{fig:JPAS_CombinedTracers_AllBins_DeWiggle_LO2_T12345_8500_CBM_Highest}, and in \cref{fig:GlobalFeature} for the 2\% global oscillatory one. The reconstruction of the global feature does not track the oscillations at large scales, but it successfully reconstructs the oscillations at the smallest scales. Although the Bayes factor is one order of magnitude smaller than that of the LO feature, and the evidences for $N > 4$ are larger, the reconstruction contours are very similar in both cases. Finally, we investigate the 1\% amplitude feature and find no statistically significant detection under any tested configuration. 

We achieved a relative error $\approx 1\%$ at the $1\sigma$ level in the reconstructions, and $\approx 4\%$ at the $3\sigma$ level. This precision is comparable to that of the Planck DR3 $P_{\mathcal{R}}(k)$ reconstructions \cite{PlanckInflation18}. Parametric forecasts for recent LSS experiments also expect to obtain few percent precision in $P_{\mathcal{R}}(k)$ (see for example \cite{EuclidSearchFeatures}).

 	\begin{table}[t]
 		\centering
 		\begin{tabular}{|c| c| c| }
 \hline
\textbf{J-PAS T12345 galaxies} & \textbf{Bayes factor} & \textbf{Hypothesis test}  \\
\hline
\multicolumn{3}{|c|}{\textbf{LO 5\%} }           \\
\hline
Blue and red, $z = 0.4$  &  $0.41$ (Negative)  &  0\% \& 15\%     \\
Blue, combined $z$-bins  &  $1.9 \times 10^{-3}$ (Decisive)  &  15\% \& 61\%     \\
Red, combined $z$-bins  &  $4.8 \times 10^{-3}$ (Decisive)  &  15\% \& 63\%     \\
\hline
\multicolumn{3}{|c|}{\textbf{LO 3\%} }           \\
\hline
Blue, combined $z$-bins   &  $1.13$ (Negative)  &  0\% \& 0\%     \\
Blue \& red, combined $z$-bins  &  $0.015$ (Very Strong)  &  17\% \& 51\%  \\
\hline
\multicolumn{3}{|c|}{\textbf{LO 2\%} }           \\
\hline
Blue \& red, combined $z$-bins  &  $0.31$ (Substantial)  &  0\% \& 21\% \\
\hline
\multicolumn{3}{|c|}{{\textbf{GO 2\%}} }           \\
\hline
Blue \& red, combined $z$-bins  &  $0.04$ (Substantial)  &  0\% \& 34\%
\\
\hline
\end{tabular}

 		\hfill 
 		\caption{Bayes factor and hypothesis test results for the reconstructions combining J-PAS $z$-bins with injected local oscillatory features of 5\%, 3\% and 2\% amplitude, and for a global feature of 2\% amplitude. The results combining blue and red galaxies provide the tightest constraints achieved in this work. The hypothesis test reports the fraction of scales exceeding the ``detection" threshold $1 - \beta > 0.9$ and above the ``hint of detection" threshold $1 - \beta > 0.5$.}
			\label{tab:CombinedBinsResults}
	\end{table}

\section{Conclusions}
\label{sec:Conclusions}

This work demonstrates the potential of J-PAS galaxy observations to constrain the primordial power spectrum, particularly in the context of detecting oscillatory features \cite{MirandaHuModeloFeature,SearchingFeatures1,SearchingFeatures2}. These features can arise in some non-standard inflationary models: local features especially in those with a step in the inflaton potential \cite{ModelAdams2001} or in the sound speed \cite{ModelAchucarro2010}, and global features arising from non-Bunch-Davies initial conditions \cite{ModelBozza2003}, axion monodromy \cite{ModelFlauger2017}, or boundary effective field theory models \cite{SearchingFeatures1, ModelJackson2013}. For the localized features the power-law behaviour is recovered at much larger ($k < 0.01 \text{ h} \text{ Mpc}^{-1}$) or smaller scales ($k > 0.5 \text{ h} \text{ Mpc}^{-1}$), which are more effectively probed by the CMB and by weak lensing or the Lyman-$\alpha$ forest, respectively. The best fits of these other probes remain largely unaltered when features are introduced in the intermediate range $k \in [0.02,0.2] \text{ h} \text{ Mpc}^{-1}$, where J-PAS provides enhanced signal-to-noise. Across this scale range, different oscillatory features---or feature templates such as bumps or linear/logarithmic oscillations---could be probed with J-PAS.

Based on this theoretical motivation, we explore how various J-PAS specifications and observational factors affect the reconstruction precision of $P_{\mathcal{R}}(k)$. For single redshift-bin reconstructions, we find that redshift-bin selection, tracer type, tray strategy, and survey area significantly influence the reconstruction performance. We select an intermediate $z = 0.4$ bin, which offers the best overall signal-to-noise ratio by balancing cosmic variance and shot noise. Among tracers, blue galaxies observed with the full tray configuration (T12345) provide the best performance at this redshift. Despite red galaxies having higher effective number densities, the higher photometric redshift precision of blue galaxies leads to more accurate reconstructions of $P_{\mathcal{R}}(k)$, consistent with the signal-to-noise analysis. The impact of both tracer type and tray strategy becomes more significant at higher redshifts, due to evolving number densities and photometric redshift errors. In the $z = 0.4$ bin, blue galaxies with the full tray strategy are preferred to the 3-tray one (T125), while red galaxies show little difference between tray strategies. Sky coverage is also a key factor in achieving high reconstruction fidelity.

At intermediate redshifts, reconstruction precision is more sensitive to photometric redshift accuracy and sky area than to number density, which mainly affects the smallest scales. Therefore, the good photometric performance and wide area coverage of J-PAS are crucial assets for high-quality reconstructions.

Consistent with this expectation, a $10\%$ amplitude feature is reliably recovered across all single-bin scenarios. Under the full J-PAS area assumption ($8500\ \text{deg}^2$), the Bayes factor indicates decisive evidence for the feature model in all cases, with the best results from the blue galaxies under T12345. Even under the half survey area ($4000 \text{ deg}^2$), the $10\%$ amplitude feature is still recovered, although with lower statistical significance. In contrast, a $5\%$ amplitude feature cannot be detected with single-bin reconstructions, even when combining both tracers.

This limitation can be overcome by combining multiple redshift bins, thereby improving the total $S/N$. Using either red or blue galaxies alone, the combined redshift bins allow decisive detection of a $5\%$ feature. Moreover, combining both tracers across redshifts enables the detection of features with amplitude as small as 2\%, with substantial statistical significance. Such levels of sensitivity are competitive with other stage IV galaxy surveys \cite{EuclidSearchFeatures} and Planck DR 3 \cite{PlanckInflation18}, and may provide key discriminatory power among non-standard inflationary scenarios.

In the present work, the only non-linear effect included is the BAO damping, which suppresses both broadband wiggles and the feature signal. We verified that this effect does not significantly alter the reconstruction results. A complete non-linear description of the galaxy power spectrum is beyond the scope of this work, as it would significantly increase the computational cost of the nested sampling. A promising compromise involves adopting parametric models of the non-linear galaxy power spectrum (see, for instance, \cite{BOSSFeatures}), which can retain essential non-linear effects at a lower computational expense. Including higher-order multipoles, mainly the quadrupole, is expected to modestly improve feature sensitivity compared to the monopole-only case. However, it could be relevant to break the degeneracy between the amplitude of the power spectrum and the galaxy bias.

Finally, the inclusion of additional J-PAS tracers, such as quasars, could further enhance the sensitivity to low-amplitude features. Preliminary $P_{\mathcal{R}}(k)$ reconstructions using miniJPAS-based estimates for quasars show poorer performance compared to galaxies, because quasars, despite their higher bias, suffer from low number densities and shot-noise-dominated power spectra. Nevertheless, these tracers probe higher redshift bins, where non-linearities have less impact, and can be used in testing feature recovery at smaller scales up to $k \approx 0.4 {\text{ h} \text{ Mpc}^{-1}}$.

\acknowledgments

The authors thank Sefa Pamuk, Juan Villafañe-Calvo, and Álvaro Álvarez-Candal for their valuable advice and comments, and Carolina Queiroz and Liliane Nakazono for their helpful clarifications regarding the J-PAS data. The authors acknowledge Santander Supercomputación support group at the University of Cantabria, who provided access to the supercomputer Altamira Supercomputer at the Institute of Physics of Cantabria (IFCA-CSIC), member of the Spanish Supercomputing Network, for performing simulations/analyses. The authors also thank the Spanish AEI and MICIU for the financial support provided under the projects with references PID2019-110610RB and PID2022-139223OB-C21, and acknowledge support from Universidad de Cantabria and Consejería de Universidades, Igualdad, Cultura y Deporte del Gobierno de Cantabria via the \textit{Instrumentación y ciencia de datos para sondear la naturaleza del universo} project, as well as from Unidad de Excelencia María de Maeztu (MDM-2017-0765). We also acknowledge financial support from the Plan Complementario AstroHEP funded by the "European Union NextGenerationEU/PRTR". \sloppy This work was supported by the MICIU (Spain) project PID2022-138263NB-I00 funded by MICIU/AEI/10.13039/501100011033 and by ERDF/EU. GMS acknowledges financial support from the Formación de Personal Investigador (FPI) programme, ref. PRE2018-085523, associated to the Spanish Agencia Estatal de Investigación (AEI, MICIU) project ESP2017-83921-C2-1-R, and also from the project UC-LIME (PID2022-140670NA-I00), financed by MCIN/AEI/ 10.13039/501100011033/FEDER, UE. RMGD acknowledges financial support from the Severo Ochoa grant CEX2021-001131-S, funded by MICIU/AEI/10.13039/501100011033, and from the project PID2022-141755NB-I00. We acknowledge the use of PolyChord \cite{MandatoryPolyChord1,MandatoryPolyChord2}, CAMB \cite{CAMB}, GetDist \cite{GetDist}, and Cobaya \cite{CobayaMandatory1,CobayaMandatory2}. The Cobaya code has been created with the help of the Python packages \texttt{NumPy} \cite{Numpy}, \texttt{Matplotlib} \cite{Matplotlib} and \texttt{SciPy} \cite{Scipy}.

Based on observations made with the JST250 telescope and JPCam at the Observatorio Astrofísico de Javalambre (OAJ), in Teruel, owned, managed, and operated by the Centro de Estudios de Física del Cosmos de Aragón (CEFCA). We acknowledge the OAJ Data Processing and Archiving Department (DPAD) for reducing and calibrating the OAJ data 
used in this work.

Funding for the J-PAS Project has been provided by the Governments of Spain and Aragón through the Fondo de Inversiones de Teruel; the Aragonese Government through the Research Groups E96, E103, E16\_17R, E16\_20R, and E16\_23R; the Spanish Ministry of Science and Innovation (MCIN/AEI/10.13039/501100011033 y FEDER, Una manera de hacer Europa) with grants PID2021-124918NB-C41, PID2021-124918NB-C42, PID2021-124918NA-C43, and PID2021-124918NB-C44; the Spanish Ministry of Science, Innovation and Universities (MCIU/AEI/FEDER, UE) with grants PGC2018-097585-B-C21 and PGC2018-097585-B-C22; the Spanish Ministry of Economy and Competitiveness (MINECO) under AYA2015-66211-C2-1-P, AYA2015-66211-C2-2, and AYA2012-30789; and European FEDER funding (FCDD10-4E-867, FCDD13-4E-2685). The Brazilian agencies FINEP, FAPESP, FAPERJ and the National Observatory of Brazil have also contributed to this project. Additional funding was provided by the Tartu Observatory and by the J-PAS Chinese Astronomical Consortium.

This paper has gone through internal review by the J-PAS collaboration.

\section*{Appendix: Impact of BAO smoothing on reconstructions}

\label{subsec:AppendixBAO}

Non-linear matter clustering damps the Baryon Acoustic Oscillations (BAOs) in the matter power spectrum. This effect can be modeled by replacing the linear matter power spectrum, $P_m(k)$, with a dewiggled version, $P_{\text{dw}}(k)$, defined as:
    \begin{equation}\label{Dewiggled}
P_{\mathrm{dw}}(k, \mu, z) \equiv P_{\mathrm{m}}(k, z) \mathrm{e}^{-g_\mu k^2}+P_{\mathrm{nw}}(k, z)\left(1-\mathrm{e}^{-g_\mu k^2}\right),
    \end{equation}
where $P_{\text{nw}}$ denotes the no-wiggle matter power spectrum. The damping scale is controlled by the function $g_{\mu}$, given by:
    \begin{equation}\label{gDewiggled}
g_\mu(\mu, z) =\left[\sigma_{v,\mathrm{fid}}(z)\right]^2\left\{1-\mu^2+\mu^2\left[1+f_{\mathrm{fid}}(z)\right]^2\right\},
    \end{equation}
where $\sigma_{v,\mathrm{fid}}$ is the variance of the displacement field, set equal to the dispersion $\sigma_{p,\mathrm{fid}}$ in \cref{FingersOfGod}.

To ensure robustness for cosmologies far from the fiducial model—such as those explored during sampling—we require a no-wiggle matter power spectrum construction that remains accurate across a wide parameter space. The Savitzky–Golay filtering method used in \cite{SavGolExample} is not well suited for this purpose. Instead, we adopt the approach described in \cite{BriedenNoWiggle}. This approach identifies all local extrema of the gradient of the rescaled power spectrum $P_m(k) \, k^{7/4}$, constructs two interpolation curves, one through maxima and another through minima, and defines the no-wiggle spectrum $P_{nw} (k,z)$ as their average.

To assess the impact of BAO smoothing, we compare the reconstructions of the primordial power spectrum $\mathcal{P}_{\mathcal{R}}(k)$ obtained using the dewiggled spectrum with those obtained using the linear matter spectrum $P_m(k)$. The resulting reconstructions show negligible differences, and the associated statistical tests yield consistent results in both cases.

\bibliographystyle{JHEP.bst}

\bibliography{bibliography}

\end{document}